\documentclass[english]{article}%
\usepackage{amssymb}
\usepackage{fontenc}
\usepackage[latin1]{inputenc}
\usepackage{amsmath}
\usepackage{fontenc}
\usepackage{babel}
\usepackage{amsfonts}
\usepackage{graphicx}%
\providecommand{\U}[1]{\protect\rule{.1in}{.1in}}
 
\begin{document}

\title{{\Large Statistical properties of states in QED with unstable vacuum}}
\author{S.P. Gavrilov\thanks{Department of Physics, Tomsk State University, Lenin ave.
36, 634050 Tomsk; Department of General and Experimental Physics, Herzen State
Pedagogical University of Russia, Moyka embankment 48, 191186 St.~Petersburg,
Russia; e-mail: gavrilovsergeyp@yahoo.com}, D.M. Gitman\thanks{Department of
Physics, Tomsk State University, Lenin ave. 36, 634050 Tomsk; P.N. Lebedev
Physical Institute, Russian Academy of Sciences, 119991 Moscow, Russia; and
Institute of Physics, University of São Paulo, CP 66318, CEP 05315-970 São
Paulo, São Paulo, Brazil; gitman@if.usp.br}, and A.A.
Shishmarev\thanks{Institute of Physics, University of São Paulo, CP 66318, CEP
05315-970 São Paulo, São Paulo, Brazil; ; e-mail: a.a.shishmarev@mail.ru}}
\maketitle

\begin{abstract}
We study statistical properties of states of massive quantized charged Dirac
and Klein-Gordon fields interacting with a background that violates the vacuum
stability, first in general terms and then for a special electromagnetic
background. As a starting point, we use a nonperturbative expression for the
density operators of such fields derived by Gavrilov \emph{et al} [S.P.
Gavrilov, D.M. Gitman, and J.L. Tomazelli, Nucl. Phys. B \textbf{795}, 645
(2008)]. We construct the reduced density operators for electron and positron
subsystems and discuss a decoherence that may occur in the course of the
evolution due to an intermediate measurement. By calculating the entropy we
study the loss of the information in QED states due to partial reductions and
a possible decoherence. Next, we consider the so-called $T$-constant external
electric field as an external background. This exactly solvable example allows
us to calculate explicitly all statistical properties of various quantum
states of the massive charged fields under consideration.



\end{abstract}

\section{Introduction}

It is known that pure states of a quantum system provide us with the maximum
possible information about this system in contrast with mixed states of the
same quantum system. A measure of the information loss of a quantum state can
be identified by the entropy of such a state. Any unitary evolution does not
change the entropy of a quantum state, and possible violations of the unitary
evolution can be registered as a change of the entropy. On the other hand, the
entanglement is an essentially quantum property associated with the quantum
non-separability of parts of a composite system. It also can be evaluated as a
specific quantum entropy. Entangled states became a powerful tool in studying
principal questions in quantum theory and in\emph{ }quantum computation and
information theories \cite{Bell,Preskill,NieCh00,BellB}. Despite a number of
publications devoted to the entropy and entanglement of quantum states, for
the related characteristics to be fully understood, more examples of various
special systems need to be considered not only in nonrelativistic quantum
mechanics, but in quantum field theory (QFT), as well. This explains recent
interest\emph{ }in studying quantum entanglement and entropy of QFT systems
with an unstable vacuum, i.e., with strong external backgrounds that may
create particles from the vacuum (see, e.g.,
\cite{LinChH10,MizEb14,Kif92,HKMP96}). In this article we would like to
attract attention to the fact that by studying QFT systems with an unstable
vacuum we create the possibility to approach problems characterized by the
loss of information, quantum entanglement, and the entropy change in a very
close relation. We present two points that explain this assertion. Let a
quantized charged Dirac or Klein-Gordon (KG) field\footnote{As basic particles
in both cases we consider electrons with the charge $q=-e$, $e>0$, whereas
their antiparticles are positrons. In the KG case both electrons and positrons
are spinless.} interact with a strong uniform external electric field. Such a
system is a QFT with an unstable vacuum, which means that the electric field
creates electron-positron pairs from the vacuum. Particle creation from the
vacuum by strong electromagnetic, Yang-Mills, and gravitational fields is a
well-known nonlinear quantum phenomenon that has many applications in modern
high energy physics. Its theoretical study has a long story that is described
in numerous works, see for example Refs.
\cite{Nik69,Haw75,BirDav82,Gitman,FroN89,GavGi96,GavGiT06}. The creation of
charged particles from the vacuum by strong electriclike fields needs
superstrong field magnitudes compared with the Schwinger critical field
$E_{\mathrm{c}}=m^{2}c^{3}/e\hbar\simeq1.3\times10^{16}\,\mathrm{V}%
\cdot\mathrm{cm}^{-1}$ \cite{schwinger}. Nevertheless, recent progress in
laser physics allows one to hope that this effect will be experimentally
observed in the near future even in laboratory conditions (see
Ref.~\cite{Dun09} for a review). Electron-hole pair creation from the vacuum
(which is an analog of the electron-positron pair creation from the vacuum)
was recently observed in graphene by its indirect influence on the graphene
conductivity \cite{Vandecasteele10} (the conductivity of graphene modified by
the particle creation was calculated in ~\cite{GavGitY12}; some other relevant
effects may be found in \cite{Katsnelson}). The particle creation effect in a
strong uniform external electric field has an additional important feature.
The external field not only creates the pairs from the vacuum, but produces
two subsystems, well separated in space: the created electrons and the\emph{
}created positrons. States of each subsystem are described by the
corresponding density matrices. Such density matrices originally were derived
in Refs. \cite{FroGi78,GavGiT06}. Here it is interesting to study the quantum
entanglement of both subsystems and its measure by\emph{ }calculating the
corresponding entropy. A change of the entropy of QFT systems with an unstable
vacuum and a quantum entanglement of the above mentioned subsystems can occur
also due to some decoherence processes. In the case under our\emph{
}consideration these might be intermediate measurements or collisions with
some semiclassical objects (e.g., with well-known impurities in the graphene).

In the present article, we study the above-mentioned characteristics, that is,
statistical properties of states of massive quantized charged Dirac or K-G
fields with a background that violates the vacuum stability, first in general
terms and then considering a specific external electromagnetic background. As
a starting point, we use a general nonperturbative expression for the density
operators of such fields derived in Ref. \cite{GavGiT06}. In Sec. \ref{S2} we
discuss such operators with different initial conditions. Reduced density
operators for electron and positron subsystems are derived in Sec. \ref{S3}.
In Sec. \ref{S4} we study a decoherence that may occur in the course of the
evolution due to an intermediate measurement and the corresponding
modifications of the complete and the reduced density operators. In Sec.
\ref{S5}, when calculating the entropy, we study the loss of information in
the QED states due to partial reductions and possible decoherence. In Sec.
\ref{S6} we consider quantized Dirac or KG fields with the so-called
$T$-constant external electric field. This exactly solvable example allows us
to calculate explicitly all statistical properties of different quantum states
of the latter massive fields. In the Appendix we briefly recall a
nonperturbative formulation of QED with strong time-dependent electric-like
background that is used in our calculations.

\section{General density operator\label{S2}}

It is convenient to introduce a generating operator $\check{R}(J)$ that allows
one to construct density operators $\check{\rho}$ with different initial
conditions (different initial states at the initial time instant
$t_{\mathrm{in}}$). This generating operator has the following form:%
\begin{align}
&  \check{R}(J)=Z^{-1}(J)\check{\underline{R}}(J),\,\mathrm{tr}\check
{R}(J)=1,\ Z(J)=\mathrm{tr}\check{\underline{R}}(J)\ ,\nonumber\\
&  \check{\underline{R}}(J)=\ :\exp\left[  \sum_{n}\left[  a_{n}^{\dagger
}(\mathrm{in})\left(  J_{n,+}-1\right)  a_{n}(\mathrm{in})+b_{n}^{\dagger
}(\mathrm{in})\left(  J_{n,-}-1\right)  b_{n}(\mathrm{in})\right]  \right]
:,\ \label{2.18}%
\end{align}
where the variables $J_{n,\zeta}$ are sources for the electron ($\zeta=+$) or
positron ($\zeta=-$) \textrm{in} operators, $Z$ is a normalization factor (the
partition function),\ and $:\cdots:$ here and in what follows means the normal
form with respect to those creation and annihilation operators that are
situated inside the colons.

Using the canonical transformation (\ref{a7}), found in the Appendix, we can
express the \textrm{in}-operators in term of the \textrm{out}-operators and
obtain $\check{R}(J)$ (and the corresponding $\check{\rho}$) in terms of the
\textrm{out} operators, \cite{GavGiT06}. Thus, $\ $%
\begin{align}
&  \check{R}(J)=Z^{-1}(J)\left\vert c_{\mathrm{v}}\right\vert ^{2}\det\left(
1+\kappa AB\right)  ^{\kappa}\check{\underline{R}}(J),\ \nonumber\\
&  \check{\underline{R}}(J)=\mathbf{:}\exp\left[  -a^{\dagger}(\mathrm{out}%
)\left(  1-D_{+}\right)  a(\mathrm{out})-b^{\dagger}(\mathrm{out})\left(
1-D_{-}\right)  b(\mathrm{out})-a^{\dagger}(\mathrm{out})C^{\dagger}%
b^{\dagger}(\mathrm{out})-b(\mathrm{out})Ca(\mathrm{out})\right]
\mathbf{:\ ,}\nonumber\\
&  D_{+}=w\left(  +|+\right)  \left(  1+\kappa AB\right)  ^{-1}\mathbb{J}%
_{+}w\left(  +|+\right)  ^{\dagger}\,,\ \ D_{-}^{T}=w\left(  -|-\right)
^{\dagger}\mathbb{J}_{-}\left(  1+\kappa BA\right)  ^{-1}w\left(  -|-\right)
\,,\ \ A\left(  J\right)  =\mathbb{J}_{+}B^{\dagger}\mathbb{J}_{-},\nonumber\\
&  C=w\left(  -|-\right)  ^{\dagger}\mathbb{J}_{-}B\left(  1+\kappa AB\right)
^{-1}\mathbb{J}_{+}w\left(  +|+\right)  ^{\dagger}+\kappa w\left(
+-|0\right)  ^{\dagger}\,,\ \ B=\kappa w\left(  0|-+\right)  \,,\text{
}\mathbb{J}_{mn,\zeta}=\delta_{mn}J_{n,\zeta},\nonumber
\end{align}
where $\kappa=+1$ for$\mathrm{\ }$the Fermi\ case and$\mathrm{\ }\kappa=-1$
for the Bose\ case. The normalization factor $Z$ has the form%
\begin{equation}
Z(J)=\exp\left\{  \kappa\sum_{n,\zeta}\left[  \ln\left(  1+\kappa J_{n,\zeta
}\right)  \right]  \right\}  =\prod_{n,\zeta}\left[  1+\kappa J_{n,\zeta
}\right]  ^{\kappa}. \label{2.20}%
\end{equation}

In what follows, we work with two important cases of the general density
operator that correspond to the initial vacuum state and to the initial
thermal state.

(a) Setting $J=0$ and using the well-known formula \cite{Ber65}, we obtain the
density operator $\check{\rho}\left(  0\right)  $ that corresponds to the
initial vacuum state
\begin{equation}
\check{\rho}(0)=\ :\exp\left\{  -\sum\limits_{n}\left[  a_{n}^{\dagger
}(\mathrm{in})\ a_{n}(\mathrm{in})\text{ }+\text{ }b_{n}^{\dagger}%
(\mathrm{in})\ b_{n}(\mathrm{in})\right]  \right\}  ^{\text{ }}%
:\ =|0,\mathrm{in}\rangle\langle0,\mathrm{in}|. \label{2.21}%
\end{equation}
From Eqs. (\ref{2.18}) we obtain this operator in terms of the \textrm{out}%
-operators%
\begin{gather}
\check{\rho}\left(  0\right)  =\ \left\vert c_{\mathrm{v}}\right\vert
^{2}\mathbf{:}\exp\left\{  -\sum\limits_{n}\left[  a_{n}^{\dagger
}(\mathrm{out})a_{n}(\mathrm{out})+b_{n}^{\dagger}(\mathrm{out})b_{n}%
(\mathrm{out})\right.  \right. \nonumber\\
\,\left.  \left.  +\kappa a_{n}^{\dagger}(\mathrm{out})w\left(  +-|0\right)
_{nn}b_{n}^{\dagger}(\mathrm{out})+\kappa b_{n}(\mathrm{out})w\left(
+-|0\right)  _{nn}^{\dagger}a_{n}(\mathrm{out})\right]  \right\}  :\ .
\label{2.22}%
\end{gather}

Differential mean numbers $N_{n,\zeta}(0|$\textrm{$in$}$)$ of \textrm{in
}electrons and positrons in the state $\check{\rho}(0)$ are zero,
\begin{equation}
N_{n,+}(0|\mathrm{in})=\mathrm{tr}\check{\rho}(0)a_{n}^{\dagger}%
(\mathrm{in})a_{n}(\mathrm{in})=0,\ \ N_{n,-}(0|\mathrm{in})=\mathrm{tr}%
\check{\rho}(0)b_{n}^{\dagger}(\mathrm{in})b_{n}(\mathrm{in})=0,\nonumber
\end{equation}
whereas differential mean numbers $N_{n,\zeta}(0|$\textrm{$out$}$)$ of
\textrm{out }electrons and positrons in the state $\check{\rho}(0)$%
\[
N_{n,+}(0|\mathrm{out})=\mathrm{tr}\text{ }\check{\rho}\left(  0\right)
a_{n}^{\dagger}(\mathrm{out})a_{n}(\mathrm{out}),\ \ N_{n,-}(0|\mathrm{out}%
)=\mathrm{tr}\text{ }\check{\rho}\left(  0\right)  b_{n}^{\dagger
}(\mathrm{out})b_{n}(\mathrm{out}),
\]
are equal and have the form
\begin{equation}
N_{n,+}(0|\mathrm{out})=N_{n,-}(0|\mathrm{out})=N_{n}(0|\mathrm{out}%
)\ ,\ N_{n}(0|\mathrm{out})=\frac{|w\left(  +-|0\right)  _{nn}|^{2}}%
{1+\kappa|w\left(  +-|0\right)  _{nn}|^{2}}, \label{2.23}%
\end{equation}

(b) To obtain the density operator $\check{\rho}(\beta)$ that corresponds to
the thermal initial state, one has to set $J_{n,\zeta}=J_{n,\zeta}\left(
\beta\right)  ,$%
\begin{equation}
J_{n,\zeta}\left(  \beta\right)  =e^{-E_{n,\zeta}},\ \,E_{n,\zeta}%
=\beta\left(  \varepsilon_{n,\zeta}-\mu_{\zeta}\right)  , \label{2.24}%
\end{equation}
where $\varepsilon_{n,\zeta}$ are energies of electrons $(\zeta=+)$ or
positrons $\left(  \zeta=-\right)  $ with quantum numbers $n$; $\mu_{\zeta}$
are the corresponding chemical potentials, and $\beta=\Theta^{-1}$, where
$\Theta$ is the absolute temperature \cite{GavGiT06}. It can be checked that
an explicit expression for $\check{\rho}(\beta)$ in terms of the \textrm{in}
operators is
\begin{align}
&  \check{\rho}(\beta)=Z_{\mathrm{gr}}^{-1}\exp\left[  -\beta\left(  \check
{H}-\sum_{\zeta}\mu_{\zeta}\check{N}_{\zeta}\right)  \right]  ,\nonumber\\
&  Z_{\mathrm{gr}}=\exp\left[  \kappa\sum_{n\zeta}\ln\left(  1+\kappa
e^{-E_{n,\zeta}}\right)  \right]  . \label{2.26}%
\end{align}
The quantity $Z_{\mathrm{gr}}$ is the partition function of the grand
canonical ensemble, $\check{H}$ is the Hamiltonian of the system (written in
terms of \textrm{in}-operators),%
\[
\check{H}=\sum_{n}\left[  a_{n}^{\dagger}(\mathrm{in})\varepsilon_{n,+}%
a_{n}(\mathrm{in})+b_{n}^{\dagger}(\mathrm{in})\varepsilon_{n,-}%
b_{n}(\mathrm{in})\right]  ,
\]
and%
\[
\check{N}_{+}=\sum_{n}\left[  a_{n}^{\dagger}(\mathrm{in})a_{n}(\mathrm{in}%
)\right]  ,\ \ \check{N}_{-}=\sum_{n}\left[  b_{n}^{\dagger}(\mathrm{in}%
)b_{n}(\mathrm{in})\right]
\]
are operators of numbers of \textrm{in}-electrons and \textrm{in}-positrons, respectively.

Let $\check{\rho}$ be the general density matrix for an arbitrary initial
state, $N_{n,\zeta}(\mathrm{\cdots}|$\textrm{$in$}$)$ be differential mean
numbers of \textrm{in }electrons or positrons in the state $\check{\rho}$, and
$N_{n,\zeta}(\mathrm{\cdots}|$\textrm{$out$}$)$ be differential mean numbers
of \textrm{out }electrons or positrons in the state $\check{\rho}$,%
\begin{align}
&  N_{n,+}(\mathrm{\cdots}|\mathrm{in})=\mathrm{tr}\text{ }\check{\rho}%
a_{n}^{\dagger}(\mathrm{in})a_{n}(\mathrm{in}),\ \ N_{n,-}(\mathrm{\cdots
}|\mathrm{in})=\mathrm{tr}\check{\rho}b_{n}^{\dagger}(\mathrm{in}%
)b_{n}(\mathrm{in}),\nonumber\\
&  N_{n,+}(\mathrm{\cdots}|\mathrm{out})=\mathrm{tr}\text{ }\check{\rho}%
a_{n}^{\dagger}(\mathrm{out})a_{n}(\mathrm{out}),\ \ N_{n,-}(\mathrm{\cdots
}|\mathrm{out})=\mathrm{tr}\check{\rho}b_{n}^{\dagger}(\mathrm{out}%
)b_{n}(\mathrm{out})\ . \label{2.30a}%
\end{align}
Calculating the traces in the \textrm{in} basis, one can see \cite{GavGiT06}
that
\begin{equation}
N_{n,\zeta}(\mathrm{\cdots}|\mathrm{out})=N_{n,\zeta}(\mathrm{\cdots
}|\mathrm{in})+N_{n}(0|\mathrm{out})\left\{  1-\kappa\left[  N_{n,+}%
(\mathrm{\cdots}|\mathrm{in})+N_{n,-}(\mathrm{\cdots}|\mathrm{in})\right]
\right\}  . \label{2.36}%
\end{equation}
In particular, differential mean numbers $N_{n,\zeta}(\beta|$\textrm{$in$}$)$
of \textrm{in-}electrons or positrons in the state $\check{\rho}(\beta)$ are
the well-known Fermi-Dirac $\left(  \kappa=+1\right)  $ or Bose-Einstein
$\left(  \kappa=-1\right)  $ distributions,
\begin{align}
N_{n,+}(\beta|\mathrm{in})  &  =\mathrm{tr}\text{ }\check{\rho}(\beta
)a_{n}^{\dagger}(\mathrm{in})a_{n}(\mathrm{in})=(e^{E_{n,+}}+\kappa
)^{-1},\nonumber\\
N_{n,-}(\beta|\mathrm{in})  &  =\mathrm{tr}\text{ }\check{\rho}(\beta
)b_{n}^{\dagger}(\mathrm{in})b_{n}(\mathrm{in})=(e^{E_{n,-}}+\kappa)^{-1}.
\label{2.29}%
\end{align}
Differential mean numbers $N_{n,\zeta}(\beta|$\textrm{$out$}$)$ of \textrm{out
}electrons or positrons in the state $\check{\rho}(\beta)$ follow immediately
from (\ref{2.36}).

\section{Reduced density operators for electron and positron
subsystems\label{S3}}

At any fixed time instant, the complete system of quantum electrons and
positrons can be conditionally divided into two subsystems: a system of
electrons and a system of positrons. Let us suppose that the external electric
field is switched off at some sufficiently long time instant $t_{\mathrm{2}}$
in such a way that at $t_{\mathrm{out}}>t_{\mathrm{2}}$ no particle creation
occurs and both subsystems are spatially separated. Thus, the particle
creation effect by the time-dependent uniform electric field provides a real
division of the complete quantum field system into the two subsystems. We can
introduce the so-called reduced density operators $\check{\rho}_{\pm}$ of the
electron subsystem and of the positron subsystem. These operators are defined
as follows:
\begin{align}
\check{\rho}_{+}  &  =\mathrm{tr}_{-}\check{\rho}=\sum_{M=0}^{\infty}%
\sum_{\{m\}}\left(  M!\right)  ^{-1}\ _{b}\langle0,\mathrm{out}|b_{m_{M}%
}(\mathrm{out})\cdots b_{m_{1}}(\mathrm{out})|\check{\rho}|b_{m_{1}}^{\dagger
}(\mathrm{out})\cdots b_{m_{M}}^{\dagger}(\mathrm{out})|0,\mathrm{out}%
\rangle_{b}\,,\nonumber\\
\check{\rho}_{-}  &  =\mathrm{tr}_{+}\check{\rho}=\sum_{M=0}^{\infty}%
\sum_{\{m\}}\left(  M!\right)  ^{-1}\ _{a}\langle0,\mathrm{out}|a_{m_{M}%
}(\mathrm{out})\cdots a_{m_{1}}(\mathrm{out})|\check{\rho}|a_{m_{1}}^{\dagger
}(\mathrm{out})\cdots a_{m_{M}}^{\dagger}(\mathrm{out})|0,\mathrm{out}%
\rangle_{a}\,, \label{3.1}%
\end{align}
where $\check{\rho}$ is the density operator of the complete system,
$|0,\mathrm{out}\rangle_{a}$ and $|0,\mathrm{out}\rangle_{b}$ are electron and
positron vacua, respectively, ($a_{m}(\mathrm{out})|0,\mathrm{out}\rangle
_{a}=0,$ $b_{m}(\mathrm{out})|0,\mathrm{out}\rangle_{b}=0,$ $|0,\mathrm{out}%
\rangle=|0,\mathrm{out}\rangle_{a}\otimes|0,\mathrm{out}\rangle_{b}$) and
$\mathrm{tr}_{\pm}$ are the so-called reduced traces. Obviously, the reduced
density operators $\check{\rho}_{\pm}$ describe mixed states.

The reduced density operators $\check{\rho}_{\pm}$ can be obtained from the
reduced generating operators $\check{R}_{\pm}\left(  J\right)  \ $which are
defined as:
\begin{equation}
\check{R}_{\pm}\left(  J\right)  =\mathrm{tr}_{\mp}\check{R}\left(  J\right)
\,. \label{3.2}%
\end{equation}
In terms of the \textrm{out}-operators these have the form%
\begin{align}
&  \check{R}_{+}\left(  J\right)  =Z_{+}^{-1}\left(  J\right)  \ \mathbf{:}%
\exp\left\{  -\sum\limits_{n}a_{n}^{\dagger}(\mathrm{out})\left[
1-K_{+}\left(  J\right)  \right]  _{nn}a_{n}(\mathrm{out})\right\}
\mathbf{:}\,,\nonumber\\
&  \check{R}_{-}\left(  J\right)  =Z_{-}^{-1}\left(  J\right)  \ \mathbf{:}%
\exp\left\{  -\sum\limits_{n}b_{n}^{\dagger}(\mathrm{out})\left[
1-K_{-}\left(  J\right)  \right]  _{nn}b_{n}(\mathrm{out})\right\}
\mathbf{:}\,,\nonumber\\
&  K_{\pm}\left(  J\right)  =D_{\pm}+C^{\dagger}\left(  1+\kappa D_{\mp}%
^{T}\right)  ^{-\kappa}C\,,\nonumber\\
&  Z_{\pm}^{-1}\left(  J\right)  =Z^{-1}\left(  J\right)  \left\vert
c_{\mathrm{v}}\right\vert ^{2}\det\left(  1+\kappa AB\right)  ^{\kappa}%
\det\left(  1+\kappa D_{\mp}\right)  ^{\kappa}\mathcal{\,}. \label{3.3}%
\end{align}
The reduced generating operators $\check{R}_{\pm}\left(  J\right)  $ allow one
to obtain the reduced density operators $\check{\rho}_{\pm}$ for different
initial states of the system. Consider as before two important cases.

(a) By setting $J=0$ in (\ref{3.3}) we obtain the reduced density operators
$\check{\rho}_{\zeta}\left(  0\right)  =\check{R}_{\zeta}\left(  0\right)  $
for both subsystems in the case when the complete system was in the vacuum
state at the initial time instant. Taking into account that%
\begin{align*}
&  K_{\pm}\left(  0\right)  =|w\left(  +-|0\right)  |^{2}=P(+-|0)P_{\mathrm{v}%
}^{-1}\,,\\
&  Z_{\pm}^{-1}\left(  0\right)  =\left\vert c_{\mathrm{v}}\right\vert
^{2}=P_{\mathrm{v}}\,,\ \ P(+-|0)=\left\vert \langle0,\mathrm{out}\left\vert
a_{n}(\mathrm{out})b_{n}(\mathrm{out})\right\vert 0,\mathrm{in}\rangle
\right\vert ^{2},
\end{align*}
where $P(+-|0)$ and $P_{\mathrm{v}}$ are probabilities of pair creation and
the vacuum-to-vacuum transition, respectively, we obtain explicit expressions
for $\check{\rho}_{\zeta}\left(  0\right)  $:
\begin{align}
\check{\rho}_{+}\left(  0\right)   &  =\check{R}_{+}\left(  0\right)
=\left\vert c_{\mathrm{v}}\right\vert ^{2}\ \mathbf{:}\exp\left\{
-\sum\limits_{n}a_{n}^{\dagger}(\mathrm{out})\left[  1-P(+-|0)P_{\mathrm{v}%
}^{-1}\right]  _{nn}a_{n}(\mathrm{out})\right\}  \mathbf{:}\ ,\nonumber\\
\check{\rho}_{-}\left(  0\right)   &  =\check{R}_{-}\left(  0\right)
=\left\vert c_{\mathrm{v}}\right\vert ^{2}\ \mathbf{:}\exp\left\{
-\sum\limits_{n}b_{n}^{\dagger}(\mathrm{out})\left[  1-P(+-|0)P_{\mathrm{v}%
}^{-1}\right]  _{nn}b_{n}(\mathrm{out})\right\}  \mathbf{:\ .} \label{3.4}%
\end{align}
It should be noted that reduced density operators (\ref{3.4}) were originally
obtained in Ref. \cite{FroGi78}.

(b) By setting the sources $J$\emph{\ }in expression (\ref{3.3})\emph{\ }%
equal\emph{ }to\emph{ }$J_{n,\zeta}\left(  \beta\right)  $\emph{\ }according
to Eqs. (\ref{2.24}), we see that the reduced generating operators (\ref{3.3})
become the reduced density operators $\check{\rho}_{\zeta}\left(
\beta\right)  $ of the system that was in thermal equilibrium at the initial
time instant $\check{R}_{+}\left(  J\right)  =\check{\rho}_{+}\left(
\beta\right)  $ and $\check{R}_{-}\left(  J\right)  =\check{\rho}_{-}\left(
\beta\right)  $.

\section{Decoherence in course of the evolution\label{S4}}

\subsection{General}

In the previous sections we considered the case where the information loss was
due to the averaging over one of the subsystems of electrons or positrons.
However, information loss can also occur due to the interaction of the quantum
system with classical (or semiclassical) objects, or, in other words, due to
decoherence. One can imagine two possible scenarios for this: It can happen
first during intermediate measurements by a classical tool and, second, as a
result of collisions of particles with some semiclassical objects (for
example, well-known impurities in the graphene). For us, there is no
difference which of the mechanisms is implemented, so in what follows we talk
about an intermediate measurement by a classical tool as a source of the decoherence.

Consider the case when the unitary evolution of the system is interrupted by a
single intermediate measurement. The external field starts to act at the time
instant $t_{\mathrm{in}}$, the system is evolving in a unitary way from
$t_{\mathrm{in}}$ to $t_{1}$, during time $T_{1}$,\ then at $t_{1}$ a
decoherence takes place, and then again the unitary evolution proceeds from
$t_{1}$ to $t_{\mathrm{out}}$ during time $T_{2}$. In this case, if we
consider the Heisenberg picture, the \textrm{out}-set of creation and
annihilation operators for electrons and positrons of the interval $T_{1}$ is
the \textrm{in} set of the interval $T_{2}$.

Suppose that during time interval $T_{1}$ the system is described by density
operator $\check{\rho}(0)$, i.e. the system is in the vacuum state at initial
time instant $t_{\mathrm{in}}$. Differential mean numbers of electrons and
positrons at the time instant $t_{1}$ are the numbers of electrons or
positrons created by the external field from the vacuum $N_{n}(0|$%
\textrm{$out$}$)$ (\ref{2.23}). The electrons and positrons created in pairs
by the external field are entangled.

During the time interval $T_{2}$ the system is described by the density
operator which we denote by $\check{\rho}^{\prime}$. The latter in terms of
the \textrm{in} set of creation-annihilation operators for electrons and
positrons must describe the system without quantum correlations between the
electrons and positrons created (i.e. new \textquotedblleft
initial\textquotedblright\ state of the system in the time interval $T_{2}$ is
the state without any entanglement).

Such an operator can be obtained by using the von Neumann reduction principle
\cite{Neuma32}. Let a system be in a pure state that is described by a state
vector $|\psi\rangle$, or equivalently by a density operator $\hat{\rho}$ that
is in such a case the projector, $\hat{\rho}=\hat{P}_{\psi}=|\psi
\rangle\langle\psi|$. In addition, let $\hat{R}$ be a self-adjoint observable
of the system. In the simplest case, when this observable has a nondegenerate
discrete spectrum the following spectral decomposition holds $\hat{R}%
=\sum_{\alpha}r_{\alpha}P_{\varphi_{\alpha}}$, where $r_{\alpha}$ are possible
eigenvalues of the observable, and $P_{\varphi_{\alpha}}$ are projectors on to
the corresponding eigenvectors $|\varphi_{\alpha}\rangle$, $\hat{P}%
_{\varphi_{\alpha}}=|\varphi_{\alpha}\rangle\langle\varphi_{\alpha}|$. When
measuring the observable $\hat{R}$, we obtain the eigenvalues $r_{\alpha}$
with the probabilities $|\langle\varphi_{\alpha}|\psi\rangle|^{2}%
=\langle\varphi_{\alpha}|\hat{P}_{\psi}|\varphi_{\alpha}\rangle=\langle
\varphi_{\alpha}|\hat{\rho}|\varphi_{\alpha}\rangle$, and just after the
measurement the state vector $|\psi\rangle$ is reduced to the vector
$|\varphi_{\alpha}\rangle$, or the density operator $\hat{\rho}$ is reduced to
the operator\textsl{ }$\hat{\rho}^{\prime}=\hat{P}_{\varphi_{\alpha}}$. A more
general case, where the system is in a mixed state, is described by the
density operator $\hat{\rho}$ with a simple discrete spectrum, $\hat{\rho
}=\sum_{n}\lambda_{n}P_{\psi_{n}}$,\ $P_{\psi_{n}}=|\psi_{n}\rangle\langle
\psi_{n}|$, $\lambda_{n}$ being statistical weights of the corresponding
states $P_{\psi_{n}}$, and $\hat{R}$ being the above-mentioned observable.Then
the measurement is presented as follows. The eigenvalues $r_{\alpha}$ are
measured with the probabilities
\[
\sum_{n}\lambda_{n}|\langle\varphi_{\alpha}|\psi_{n}\rangle|^{2}%
=\langle\varphi_{\alpha}|\hat{\rho}|\varphi_{\alpha}\rangle,
\]
and just after the measurement the density operator $\hat{\rho}$ is reduced to
the operator $\hat{\rho}^{\prime}$,
\[
\hat{\rho}^{\prime}=\sum_{\alpha}\langle\varphi_{\alpha}|\hat{\rho}%
|\varphi_{\alpha}\rangle\hat{P}_{\varphi_{\alpha}}\ .
\]

The density operator $\check{\rho}(0)$ is
\begin{equation}
\check{\rho}(0)=|0,\mathrm{in}\rangle\langle0,\mathrm{in}|. \label{4.1}%
\end{equation}
The \textrm{in} vacuum $|0,\mathrm{in}\rangle$ is connected to the
\textrm{out}-vacuum $|0,\mathrm{out}\rangle$ by relation (\ref{a8}). Then
density operator $\check{\rho}(0)$ can be presented as
\begin{equation}
\check{\rho}(0)=V|0,\mathrm{out}\rangle\langle0,\mathrm{out}|V^{\dagger}.
\label{4.2}%
\end{equation}
We are interested in the case of a uniform external field, which does not mix
different quantum modes. Then amplitudes (\ref{a6}) are diagonal. Thus, it is
possible to factorize $V$ defined by\ (\ref{a8}) as%
\begin{equation}
V=\prod\limits_{n}V_{n},\text{ \ }V_{n}=v_{4n}v_{3n}v_{2n}v_{1n}, \label{4.3}%
\end{equation}
where
\begin{align}
&  v_{1n}=\exp\left\{  -\kappa b_{n}(\mathrm{out})w\left(  0|-+\right)
_{nn}a_{n}(\mathrm{out})\right\}  \,,\;v_{2n}=\exp\left\{  a_{n}^{\dagger
}(\mathrm{out})\left[  \ln w\left(  +|+\right)  \right]  _{nn}a_{n}%
(\mathrm{out})\right\}  \,,\nonumber\\
&  v_{3n}=\exp\left\{  -\kappa b_{n}(\mathrm{out})\left[  \ln w\left(
-|-\right)  \right]  _{nn}b_{n}^{\dagger}(\mathrm{out})\right\}
\,,\;v_{4n}=\exp\left\{  -\kappa a_{n}^{\dagger}(\mathrm{out})w\left(
+-|0\right)  _{nn}b_{n}^{\dagger}(\mathrm{out})\right\}  \,.\nonumber
\end{align}

Making use of the explicit form of $V_{n}$ and definition (\ref{a8}) we can
write
\begin{equation}
|0,\mathrm{in}\rangle=c_{\mathrm{v}}\prod\limits_{n}\sum_{m=0}\frac{\left(
-1\right)  ^{m}}{m!}\left[  \kappa w\left(  +-|0\right)  _{nn}a_{n}^{\dagger
}(\mathrm{out})b_{n}^{\dagger}(\mathrm{out})\right]  ^{m}|0,\mathrm{out}%
\rangle, \label{4.4a}%
\end{equation}
and then we can easily calculate $c_{\mathrm{v}}$,%
\begin{equation}
c_{\mathrm{v}}=\prod\limits_{n}\left[  w\left(  -|-\right)  _{nn}\right]
^{-\kappa}. \label{4.4}%
\end{equation}
Thus, density operators $\check{\rho}(0)$ (\ref{4.2}) can be represented as%
\begin{align}
\  &  \check{\rho}(0)=|c_{\mathrm{v}}|^{2}\prod\limits_{n}\left(
\sum\limits_{m=0}^{\infty}\frac{\left[  -\kappa w\left(  +-|0\right)
_{nn}a_{n}^{\dagger}(\mathrm{out})b_{n}^{\dagger}(\mathrm{out})\right]  ^{m}%
}{m!}\right) \nonumber\\
&  \times\check{P}_{0}\prod\limits_{n^{\prime}}\left(  \sum\limits_{m^{\prime
}=0}^{\infty}\frac{\left[  -\kappa w\left(  +-|0\right)  _{n^{\prime}%
n^{\prime}}^{\dagger}b_{n^{\prime}}(\mathrm{out})a_{n^{\prime}}(\mathrm{out}%
)\right]  ^{m^{\prime}}}{m^{\prime}!}\right)  , \label{4.5}%
\end{align}
where $\check{P}_{0}=|0,\mathrm{out}\rangle\langle0,\mathrm{out}|$.

\subsection{Measurement of differential mean numbers in the system}

Suppose that we are going to measure the physical quantity, which is the
number of particles, in the state $\check{\rho}(0)$ of the system under
consideration. The operator corresponding to such physical quantity is
$\check{N}(\mathrm{out})$,%
\begin{equation}
\check{N}(\mathrm{out})=\sum\limits_{n,\zeta}\check{N}_{n,\zeta}%
(\mathrm{out})=\sum\limits_{n}\left[  a_{n}^{\dagger}(\mathrm{out}%
)a_{n}(\mathrm{out})+b_{n}^{\dagger}(\mathrm{out})b_{n}(\mathrm{out})\right]
. \label{4.6}%
\end{equation}
Its eigenstates are mutually orthonormal vectors of the form \
\begin{align*}
&  \ |s,\mathrm{out}\rangle=|\{i,l\}_{LP},\mathrm{out}\rangle_{a}%
\otimes|\{j,k\}_{KQ},\mathrm{out}\rangle_{b},\ \\
&  \ |\{i,l\}_{LP},\mathrm{out}\rangle_{a}=\frac{\left[  a_{i_{1}}^{\dagger
}(\mathrm{out})\right]  ^{l_{1}}}{\sqrt{l_{1}!}}\cdots\frac{\left[  a_{i_{P}%
}^{\dagger}(\mathrm{out})\right]  ^{l_{P}}}{\sqrt{l_{P}!}}|0,\mathrm{out}%
\rangle_{a},\\
&  \ |\{j,k\}_{KQ},\mathrm{out}\rangle_{b}=\frac{\left[  b_{j_{1}}^{\dagger
}(\mathrm{out})\right]  ^{k_{1}}}{\sqrt{k_{1}!}}\cdots\frac{\left[  b_{j_{Q}%
}^{\dagger}(\mathrm{out})\right]  ^{k_{Q}}}{\sqrt{k_{Q}!}}|0,\mathrm{out}%
\rangle_{b},\\
&  L=0,1,2,\ldots,\text{\ \ }P=1,2,\ldots L,\text{ \ }i=i_{1},\ldots
,i_{P},\text{ \ }l_{1}+l_{2}+\ldots+l_{P}=L,\\
&  K=0,1,2,\ldots,\text{\ \ }Q=1,2,\ldots K,\text{ \ }j=j_{1},\ldots
,j_{Q},\text{ \ }k_{1}+k_{2}+\ldots+k_{Q}=K,
\end{align*}
such that its eigenvalues are%
\[
\hat{N}(\mathrm{out})|s,\mathrm{out}\rangle=\left(  L+K\right)
|s,\mathrm{out}\rangle,
\]
where $s$ is the full set of quantum numbers $K$, $L$, $\{i\}$, $\{j\}$, $P$,
and $Q$ and $|\{i,l\}_{LP},\mathrm{out}\rangle_{a}$ is a state with $L$
electrons distributed in $P$ groups $i_{1},\ldots,i_{P}$, with $l_{1}$
electrons in the group $i_{1}$, $l_{2}$ electrons in the group $i_{2}$, and so
on. Analogously, $|\{j,k\}_{KQ},\mathrm{out}\rangle_{b}$ is a state with $K$
positrons distributed in $Q$ groups $j_{1},\ldots,j_{Q}$, with $k_{1}$
positrons in group $j_{1}$, $k_{2}$ positrons in group $j_{2}$, and so on.

According to von Neumann \cite{Neuma32}, the density operator $\check{\rho
}(0)$ is after such measurement reduced to the operator $\check{\rho}_{N}$ of
the form%
\begin{equation}
\check{\rho}_{N}=\sum\limits_{s}\langle s,\mathrm{out}|\check{\rho
}(0)|s,\mathrm{out}\rangle\check{P}_{s},\text{ \ }\check{P}_{s}%
=|s,\mathrm{out}\rangle\langle s,\mathrm{out}|. \label{4.8}%
\end{equation}
Due to the structure of the density operator $\check{\rho}(0)$ given by
Eq.~(\ref{4.5})$,$ the weights $\langle s,\mathrm{out}|\check{\rho
}(0)|s,\mathrm{out}\rangle$ are non-zero only when the states $|s,\mathrm{out}%
\rangle$ are states with an integer number of pairs. Thus, we obtain%
\begin{align}
&  \check{\rho}_{N}=|c_{\mathrm{v}}|^{2}\sum\limits_{f}W_{f}\check{P}%
_{f},\text{ \ }\sum\limits_{f}=\sum\limits_{M=0}^{\infty}\sum\limits_{Z=1}^{M}%
{\displaystyle\sum\limits_{\{m,n\}}}
,\text{ \ }\check{P}_{f}=|f,\mathrm{out}\rangle\langle f,\mathrm{out}%
|,\nonumber\\
&  W_{f}=|w\left(  +-|0\right)  _{n_{1}n_{1}}|^{2m_{1}}\ldots|w\left(
+-|0\right)  _{n_{Z}n_{Z}}|^{2m_{Z}},\text{ \ }m_{1}+m_{2}+\ldots
+m_{z}=M,\nonumber\\
&  \ |f,\mathrm{out}\rangle=\frac{\left[  a_{n_{1}}^{\dagger}(\mathrm{out}%
)b_{n_{1}}^{\dagger}(\mathrm{out})\right]  ^{m_{1}}}{m_{1}!}\cdots
\frac{\left[  a_{n_{Z}}^{\dagger}(\mathrm{out})b_{n_{Z}}^{\dagger
}(\mathrm{out})\right]  ^{m_{Z}}}{m_{Z}!}|0,\mathrm{out}\rangle, \label{4.8b}%
\end{align}
where $f$ is a complete set of quantum numbers $M$, $Z$, $\{m\}$, and $\{n\}$
and $|f,\mathrm{out}\rangle$ is a state with the total number of pairs $M$
distributed in $Z$ groups, $m_{1}$ pairs being in the group $n_{1}$, $m_{2}$
pairs being in the group $n_{2}$ and so on. Unlike (\ref{4.5}), the latter
expression contains only terms diagonal in $f$. Thus, the measurement destroy
nondiagonal terms of the density operator (\ref{4.5}).

Let us now calculate the reduced (in the sense of Sec. \ref{S3}) operators
$\left[  \check{\rho}_{N}\right]  _{\zeta}$:%
\begin{align}
\left[  \check{\rho}_{N}\right]  _{\zeta}  &  =\mathrm{tr}_{-\zeta}\text{
}\check{\rho}_{N}=|c_{\mathrm{v}}|^{2}\sum\limits_{f}W_{f}\text{ }%
\mathrm{tr}_{-\zeta}\text{ }\check{P}_{f}\ ,\nonumber\\
\mathrm{tr}_{-}\text{ }\check{P}_{f}  &  =\frac{\left[  a_{n_{1}}^{\dagger
}(\mathrm{out})\right]  ^{m_{1}}}{\sqrt{m_{1}!}}\cdots\frac{\left[  a_{n_{Z}%
}^{\dagger}(\mathrm{out})\right]  ^{m_{Z}}}{\sqrt{m_{Z}!}}|0,\mathrm{out}%
\rangle_{aa}\langle0,\mathrm{out}|\frac{\left[  a_{n_{Z}}(\mathrm{out}%
)\right]  ^{m_{Z}}}{\sqrt{m_{Z}!}}\cdots\frac{\left[  a_{n_{1}}(\mathrm{out}%
)\right]  ^{m_{1}}}{\sqrt{m_{1}!}},\nonumber\\
\mathrm{tr}_{+}\text{ }P_{f}  &  =\frac{\left[  b_{n_{1}}^{\dagger
}(\mathrm{out})\right]  ^{m_{1}}}{\sqrt{m_{1}!}}\cdots\frac{\left[  b_{n_{Z}%
}^{\dagger}(\mathrm{out})\right]  ^{m_{Z}}}{\sqrt{m_{Z}!}}|0,\mathrm{out}%
\rangle_{bb}\langle0,\mathrm{out}|\frac{\left[  b_{n_{Z}}(\mathrm{out}%
)\right]  ^{m_{Z}}}{\sqrt{m_{Z}!}}\cdots\frac{\left[  b_{n_{1}}(\mathrm{out}%
)\right]  ^{m_{1}}}{\sqrt{m_{1}!}}. \label{4.10}%
\end{align}
On the other hand, one can calculate the reduced density operators
$\check{\rho}_{\zeta}(0)$ by taking reduced traces (\ref{3.1}) of the operator
(\ref{4.5}) to verify that they have exactly the same form,
\[
\left[  \check{\rho}_{N}\right]  _{\zeta}=\check{\rho}_{\zeta}(0).
\]

\subsection{Measurements of differential mean numbers in the subsystems}

Suppose now that we measure the number of either electrons or positrons. The
corresponding operators of these physical quantities are%
\begin{align}
\check{N}_{+}(\mathrm{out})  &  =\sum\limits_{n}\check{N}_{n,+}(\mathrm{out}%
)=\sum\limits_{n}a_{n}^{\dagger}(\mathrm{out})a_{n}(\mathrm{out}),\nonumber\\
\check{N}_{-}(\mathrm{out})  &  =\sum\limits_{n}\check{N}_{n,-}(\mathrm{out}%
)=\sum\limits_{n}b_{n}^{\dagger}(\mathrm{out})b_{n}(\mathrm{out}).
\label{4.12}%
\end{align}
The spectra of the operators (\ref{4.12}) are%
\begin{align*}
&  \ |s_{+},\mathrm{out}\rangle_{a}=|\{i,l\}_{LP},\mathrm{out}\rangle
_{a}=\frac{\left[  a_{i_{1}}^{\dagger}(\mathrm{out})\right]  ^{l_{1}}}%
{\sqrt{l_{1}!}}\cdots\frac{\left[  a_{i_{P}}^{\dagger}(\mathrm{out})\right]
^{l_{P}}}{\sqrt{l_{P}!}}|0,\mathrm{out}\rangle_{a},\\
&  L=0,1,2,\ldots,\text{\ \ }P=1,2,\ldots L,\text{ \ }i=i_{1},\ldots
,i_{P},\text{ \ }l_{1}+l_{2}+\ldots+l_{P}=L,\\
&  \ |s_{-},\mathrm{out}\rangle_{b}=|\{j,k\}_{KQ},\mathrm{out}\rangle
_{b}=\frac{\left[  b_{j_{1}}^{\dagger}(\mathrm{out})\right]  ^{k_{1}}}%
{\sqrt{k_{1}!}}\cdots\frac{\left[  b_{j_{Q}}^{\dagger}(\mathrm{out})\right]
^{k_{Q}}}{\sqrt{k_{Q}!}}|0,\mathrm{out}\rangle_{b},\\
&  K=0,1,2,\ldots,\text{\ \ }Q=1,2,\ldots K,\text{ \ }j=j_{1},\ldots
,j_{Q},\text{ \ }k_{1}+k_{2}+\ldots+k_{Q}=K,\\
&  \hat{N}_{+}(\mathrm{out})|s_{+},\mathrm{out}\rangle_{a}=L|s_{+}%
,\mathrm{out}\rangle_{a},\ \ \hat{N}_{-}(\mathrm{out})|s_{-},\mathrm{out}%
\rangle_{b}=K|s_{-},\mathrm{out}\rangle_{b}\ .
\end{align*}
The states $|\{i,l\}_{LP},\mathrm{out}\rangle_{a}$ and $|\{j,k\}_{KQ}%
,\mathrm{out}\rangle_{b}$ are defined in the same way as in the previous section.

The density operators after such measurements, which we denote by $\check
{\rho}_{N_{+}}$ and $\check{\rho}_{N_{-}}$, respectively, have the form%
\begin{align}
\check{\rho}_{N_{+}}  &  =\sum\limits_{s+}\text{ }_{a}\langle s_{+}%
,\mathrm{out}|\check{\rho}(0)|s_{+},\mathrm{out}\rangle_{a}P_{s+},\text{
\ }P_{s+}=|s_{+},\mathrm{out}\rangle_{aa}\langle s_{+},\mathrm{out}|,\text{
\ }\sum\limits_{s+}=\sum\limits_{L=0}^{\infty}\sum\limits_{P=1}^{L}%
\sum\limits_{\{i,l\}}\ ,\nonumber\\
\check{\rho}_{N_{-}}  &  =\sum\limits_{s-}\text{ }_{b}\langle s_{-}%
,\mathrm{out}|\check{\rho}(0)|s_{-},\mathrm{out}\rangle_{b}P_{s-},\text{
\ }P_{s-}=|s_{-},\mathrm{out}\rangle_{bb}\langle s_{-},\mathrm{out}|,\text{
\ }\sum\limits_{s-}=\sum\limits_{K=0}^{\infty}\sum\limits_{Q=1}^{K}%
\sum\limits_{\{j,k\}}\ . \label{4.15}%
\end{align}

Let us now calculate the quantities $_{a}\langle s_{+},\mathrm{out}%
|\check{\rho}(0)|s_{+},\mathrm{out}\rangle_{a}P_{s+}\,$and $_{b}\langle
s_{-},\mathrm{out}|\check{\rho}(0)|s_{-},\mathrm{out}\rangle_{b}P_{s-}$. Due
to the structure of $\check{\rho}(0)$ they are equal and have the form%
\begin{align}
&  \ _{a}\langle s_{+},\mathrm{out}|\check{\rho}(0)|s_{+},\mathrm{out}%
\rangle_{a}P_{s_{+}}=_{b}\langle s_{-},\mathrm{out}|\check{\rho}%
(0)|s_{-},\mathrm{out}\rangle_{b}P_{s_{-}}\nonumber\\
&  \ =|c_{\mathrm{v}}|^{2}\frac{\left[  -\kappa w\left(  +-|0\right)
_{i_{1}i_{1}}a_{i_{1}}^{\dagger}(\mathrm{out})b_{i_{1}}^{\dagger}%
(\mathrm{out})\right]  ^{l_{1}}}{l_{1}!}\cdots\frac{\left[  -\kappa w\left(
+-|0\right)  _{i_{p}i_{p}}a_{i_{p}}^{\dagger}(\mathrm{out})b_{i_{p}}^{\dagger
}(\mathrm{out})\right]  ^{l_{p}}}{l_{p}!}|0,\mathrm{out}\rangle\nonumber\\
&  \ \times\langle0,\mathrm{out}|\frac{\left[  -\kappa w\left(  +-|0\right)
_{i_{p}i_{p}}^{\dagger}b_{i_{p}}(\mathrm{out})a_{i_{p}}(\mathrm{out})\right]
^{l_{p}}}{l_{p}!}\cdots\frac{\left[  -\kappa w\left(  +-|0\right)
_{i_{1}i_{1}}^{\dagger}b_{i_{1}}(\mathrm{out})a_{i_{1}}(\mathrm{out})\right]
^{l_{1}}}{l_{1}!}. \label{4.16}%
\end{align}

It is not difficult to see that density operators $\check{\rho}_{N_{+}}$ and
$\check{\rho}_{N-}$ have exactly the same form as in (\ref{4.8b}), namely,
they are sums over all possible projectors on states with an integer number of
pairs
\begin{equation}
\check{\rho}_{N+}=\check{\rho}_{N-}=\check{\rho}_{N}. \label{4.17}%
\end{equation}

Thus, we stress that measurements of $N$, $N_{+}$ and $N_{-}$ produce the same
reductions. The reduced density operators $\left[  \check{\rho}_{N+}\right]
_{\zeta}=\mathrm{tr}_{-\zeta}$ $\check{\rho}_{N+}$ and $\left[  \check{\rho
}_{N-}\right]  _{\zeta}=\mathrm{tr}_{-\zeta}$ $\check{\rho}_{N+}$ are equal to
the reduced density operators $\check{\rho}_{\zeta}(0)$ given in (\ref{3.4}).

It is also an interesting task to consider the case when the unitary evolution
of the system is interrupted by multiple measurements. However, because there
are significant technical difficulties, this problem is not considered in this paper.

\section{Entropy and entanglement of electron and positron
subsystems\label{S5}}

As already said in the Introduction, the measure of the information loss in a
quantum state $\check{\rho}$ can be identified with the entropy of such a
state, namely, with the von Neumann information entropy $S$ \cite{Neuma32},%
\begin{equation}
S\left(  \check{\rho}\right)  =-k_{B}\ \mathrm{tr}\check{\rho}\ln\check{\rho
}\ . \label{5.1}%
\end{equation}

Let $\hat{\rho}\left(  t_{\mathrm{in}}\right)  =\check{\rho}\left(
\beta\right)  $, where $\check{\rho}\left(  \beta\right)  $ is given by
(\ref{2.26}), then
\begin{equation}
S\left(  \check{\rho}\left(  \beta\right)  \right)  =k_{B}\left[  \ln
Z_{gr}+\sum_{n\zeta}E_{n,\zeta}N_{n,\zeta}(\beta|\mathrm{in})\right]  ,
\label{5.2}%
\end{equation}
The corresponding differential mean numbers $N_{n,\zeta}(\beta|$\textrm{$in$%
}$)$ are Fermi-Dirac or Bose-Einstein distributions, given by (\ref{2.29}).
The entropy (\ref{5.2}) can be written in terms of the Bose (Fermi) occupation
number alone, if we take into account that
\begin{equation}
e^{-E_{n,\zeta}}=\frac{N_{n,\zeta}(\beta|\mathrm{in})}{1-\kappa N_{n,\zeta
}(\beta|\mathrm{in})}.\text{\ } \label{5.3}%
\end{equation}
Then%
\begin{equation}
S\left(  \check{\rho}\left(  \beta\right)  \right)  =-k_{B}\sum_{n\zeta
}\left\{  \kappa\left[  1-\kappa N_{n,\zeta}(\beta|\mathrm{in})\right]
\ln\left[  1-\kappa N_{n,\zeta}(\beta|\mathrm{in})\right]  +N_{n,\zeta}%
(\beta|\mathrm{in})\ln N_{n,\zeta}(\beta|\mathrm{in})\right\}  . \label{5.4}%
\end{equation}
This expression has a form similar to expressions for entropy of the grand
canonical ensemble for Fermi- and Bose-particles \cite{Landau}.

Especially interesting information is obtained by calculating the von Neumann
information entropy of the reduced density operators of both the electron and
positron subsystems $S\left(  \hat{\rho}_{\pm}\right)  $,
\begin{equation}
S\left(  \hat{\rho}_{\pm}\right)  =-k_{B}\mathrm{tr}_{\pm}(\hat{\rho}_{\pm}%
\ln\hat{\rho}_{\pm}). \label{5.5}%
\end{equation}
According to the general theory they coincide $S\left(  \hat{\rho}_{+}\right)
=S\left(  \hat{\rho}_{-}\right)  $ and can be treated as a measure of the
quantum entanglement of these subsystems.

It is also known that one can recognize entanglement by evaluating the
so-called Schmidt measure, which is the trace of the squared reduced density
operators \cite{EisBr01}%
\begin{equation}
\tilde{S}\left(  \hat{\rho}_{\pm}\right)  =-\mathrm{tr}\left[  \left(
\hat{\rho}_{\pm}\right)  ^{2}\right]  . \label{5.6}%
\end{equation}

Let us calculate the entropy for both the electron and positron subsystems in
two important cases of the vacuum initial state and the thermal initial state
that are described by the reduced density operators $\check{\rho}_{\zeta}(0)$
and $\check{\rho}_{\zeta}(\beta)$.

\subsection{Vacuum initial state}

The entropy for the reduced density operator of the system with an initial
vacuum state has the form
\begin{equation}
S\left(  \check{\rho}_{\zeta}(0)\right)  =-k_{B}\mathrm{tr}_{\zeta}\left(
\check{\rho}_{\zeta}(0)\ln\check{\rho}_{\zeta}(0)\right)  . \label{5.7}%
\end{equation}
The term $\ln\check{\rho}_{\zeta}(0)$ on the right-hand side of (\ref{5.7})
can be written as%
\begin{align}
\ln\check{\rho}_{+}(0)  &  =\ln\left[  \left\vert c_{\mathrm{v}}\right\vert
^{2}\ \mathbf{:}\exp\left\{  -\sum_{n}a_{n}^{\dagger}(\mathrm{out})\left(
1-P(+-|0)P_{\mathrm{v}}^{-1}\right)  _{nn}a_{n}(\mathrm{out})\right\}
\mathbf{:}\right]  ,\nonumber\\
\ln\check{\rho}_{-}(0)  &  =\ln\left[  \left\vert c_{\mathrm{v}}\right\vert
^{2}\ \mathbf{:}\exp\left\{  -\sum_{n}b_{n}^{\dagger}(\mathrm{out})\left(
1-P(+-|0)P_{\mathrm{v}}^{-1}\right)  _{nn}b_{n}(\mathrm{out})\right\}
\mathbf{:}\right]  . \label{5.8}%
\end{align}
Transforming the normal-form exponents into ordinary exponents (see, for
example, \cite{GavGiT06}) and recalling that $\left\vert c_{\mathrm{v}%
}\right\vert ^{2}=P_{\mathrm{v}}$, we obtain%
\begin{align}
\ln\check{\rho}_{+}(0)  &  =\ln P_{\mathrm{v}}+\sum_{n}a_{n}^{\dagger
}(\mathrm{out})\ln\left[  P(+-|0)P_{\mathrm{v}}^{-1}\right]  _{nn}%
a_{n}(\mathrm{out}),\nonumber\\
\ln\check{\rho}_{-}(0)  &  =\ln P_{\mathrm{v}}+\sum_{n}b_{n}^{\dagger
}(\mathrm{out})\ln\left[  P(+-|0)P_{\mathrm{v}}^{-1}\right]  _{nn}%
b_{n}(\mathrm{out}). \label{5.9}%
\end{align}
Taking into account that the matrices $P(+-|0)P_{\mathrm{v}}^{-1}$ are
diagonal, one can rewrite (\ref{5.7}) as
\begin{gather}
S\left(  \check{\rho}_{+}(0)\right)  =-k_{B}\left\{  \ln P_{\mathrm{v}}%
+\sum_{n}\mathrm{tr}_{+}\left(  \check{\rho}_{+}(0)a_{n}^{\dagger
}(\mathrm{out})a_{n}(\mathrm{out})\right)  \ln\left[  P(+-|0)P_{\mathrm{v}%
}^{-1}\right]  _{nn}\right\}  ,\nonumber\\
\text{ }S\left(  \check{\rho}_{-}(0)\right)  =-k_{B}\left\{  \ln
P_{\mathrm{v}}+\sum_{n}\mathrm{tr}_{-}\left(  \check{\rho}_{-}(0)b_{n}%
^{\dagger}(\mathrm{out})b_{n}(\mathrm{out})\right)  \ln\left[
P(+-|0)P_{\mathrm{v}}^{-1}\right]  _{nn}\right\}  , \label{5.10}%
\end{gather}
where $\mathrm{tr}_{+}$ $\check{\rho}_{+}(0)a_{n}^{\dagger}(\mathrm{out}%
)a_{n}(\mathrm{out})=N_{n}(0|$\textrm{$out$}$)$ and \textrm{tr}$_{-}$
$\check{\rho}_{-}(0)b_{n}^{\dagger}(\mathrm{out})b_{n}(\mathrm{out})=N_{n}%
(0|$\textrm{$out$}$)$ are differential mean numbers of \textrm{out}-electrons
and \textrm{out}-positrons, respectively. They obviously coincide. Thus, the
entropy takes the form
\begin{equation}
S\left(  \check{\rho}_{\zeta}(0)\right)  =-k_{B}\left\{  \ln P_{\mathrm{v}%
}+\sum_{n}N_{n}(0|\mathrm{out})\left[  \ln P(+-|0)P_{\mathrm{v}}^{-1}\right]
_{nn}\right\}  . \label{5.11}%
\end{equation}

One can use the pair creation probability and the vacuum-to-vacuum probability
written in terms of differential mean numbers (see, for example,
\cite{GavGi96})
\begin{equation}
P(-+|0)_{n,n^{\prime}}=\delta_{n,n^{\prime}}\frac{P_{\mathrm{v}}%
N_{n}(0|\mathrm{out})}{1-\kappa N_{n}(0|\mathrm{out})},\quad P_{\mathrm{v}%
}=\exp\left\{  \kappa\sum_{n}\ln\left[  1-\kappa N_{n}(0|\mathrm{out})\right]
\right\}  , \label{5.12}%
\end{equation}
to obtain%
\begin{align}
&  S\left(  \check{\rho}_{\zeta}(0)\right)  =\sum_{n}S(\check{\rho}_{n,\zeta
}(0)),\ \ \nonumber\\
&  \ S(\check{\rho}_{n,\zeta}(0))=-k_{B}\left\{  \kappa\left[  1-\kappa
N_{n}\left(  \mathrm{0}|\mathrm{out}\right)  \right]  \ln\left[  1-\kappa
N_{n}\left(  \mathrm{0}|\mathrm{out}\right)  \right]  +N_{n}\left(
\mathrm{0}|\mathrm{out}\right)  \ln N_{n}\left(  \mathrm{0}|\mathrm{out}%
\right)  \right\}  . \label{5.13}%
\end{align}

Let us consider the Schmidt entanglement measure (\ref{5.6}),
\begin{equation}
\tilde{S}(\check{\rho}_{\zeta}(0))=-\mathrm{tr}\text{ }\left[  \check{\rho
}_{\zeta}(0)\right]  ^{2}. \label{5.14}%
\end{equation}
Here
\begin{align}
\left[  \check{\rho}_{+}(0)\right]  ^{2}  &  =P_{\mathrm{v}}^{2}\left\{
\mathbf{:}\exp\left[  -\sum_{n}a_{n}^{\dagger}(\mathrm{out})\left(
1-P(+-|0)P_{\mathrm{v}}^{-1}\right)  _{nn}a_{n}(\mathrm{out})\right]
\mathbf{:}\right\}  ^{2}\mathbf{,}\nonumber\\
\left[  \check{\rho}_{-}(0)\right]  ^{2}  &  =P_{\mathrm{v}}^{2}\left\{
\mathbf{:}\exp\left[  -\sum_{n}b_{n}^{\dagger}(\mathrm{out})\left(
1-P(+-|0)P_{\mathrm{v}}^{-1}\right)  _{nn}b_{n}(\mathrm{out})\right]
\mathbf{:}\right\}  ^{2}\mathbf{.} \label{5.15}%
\end{align}
Using the relation (here $D$ and $\tilde{D}$ are matrices)%
\begin{equation}
\mathbf{:}\exp\left[  -a^{\dagger}(\mathrm{out})Da(\mathrm{out})\right]
\mathbf{::}\exp\left[  -a^{\dagger}(\mathrm{out})\tilde{D}a(\mathrm{out}%
)\right]  \mathbf{:=:}\exp\left[  -a^{\dagger}(\mathrm{out})\left(
D+\tilde{D}-D\tilde{D}\right)  a(\mathrm{out})\right]  \mathbf{:\ ,}
\label{5.16}%
\end{equation}
we obtain%
\begin{align}
\left[  \check{\rho}_{+}(0)\right]  ^{2}  &  =P_{\mathrm{v}}^{2}\mathbf{:}%
\exp\left\{  \sum_{n}a_{n}^{\dagger}(\mathrm{out})\left[  \left(
P(+-|0)P_{\mathrm{v}}^{-1}\right)  ^{2}-1\right]  _{nn}a_{n}(\mathrm{out}%
)\right\}  \mathbf{:\ ,}\nonumber\\
\left[  \check{\rho}_{-}(0)\right]  ^{2}  &  =P_{\mathrm{v}}^{2}\mathbf{:}%
\exp\left\{  \sum_{n}b_{n}^{\dagger}(\mathrm{out})\left[  \left(
P(+-|0)P_{\mathrm{v}}^{-1}\right)  ^{2}-1\right]  _{nn}b_{n}(\mathrm{out}%
)\right\}  \mathbf{:\ .} \label{5.17}%
\end{align}
Calculating the traces in (\ref{5.14}) taking into account (\ref{5.12}), we
finally obtain%
\begin{align}
&  \ \tilde{S}(\check{\rho}_{\zeta}(0))=-P_{\mathrm{v}}^{2}\det\left[
1+\kappa\left(  P(+-|0)P_{\mathrm{v}}^{-1}\right)  ^{2}\right]  ^{\kappa
}\nonumber\\
&  =-\prod_{n}\left\{  1-2\kappa N_{n}(0|\mathrm{out})+(1+\kappa)\left[
N_{n}(0|\mathrm{out})\right]  ^{2}\right\}  ^{\kappa}. \label{5.18}%
\end{align}

\subsection{Thermal initial state}

The entropy for the operators $\check{\rho}_{\zeta}(\beta)$, which describe
the system that has been in thermal equilibrium at the initial time instant,
has the form%
\begin{equation}
S\left(  \check{\rho}_{\beta,\zeta}\right)  =-k_{B}\mathrm{tr}_{\zeta}\text{
}\check{\rho}_{\zeta}(\beta)\ln\check{\rho}_{\zeta}(\beta).\label{5.19}%
\end{equation}
Transforming the expressions $\ln\check{\rho}_{\zeta}(\beta)$ as
\begin{align}
\ln\check{\rho}_{+}(\beta) &  =\ln Z_{\zeta}(J_{\beta})+\sum_{n}a_{n}%
^{\dagger}(\mathrm{out})\ln\left[  K_{+}(J_{\beta})\right]  _{nn}%
a_{n}(\mathrm{out}),\nonumber\\
\ln\check{\rho}_{-}(\beta) &  =\ln Z_{\zeta}(J_{\beta})+\sum_{n}b_{n}%
^{\dagger}(\mathrm{out})\ln\left[  K_{-}(J_{\beta})\right]  _{nn}%
b_{n}(\mathrm{out}),\label{5.20}%
\end{align}
one can write
\begin{equation}
S\left(  \check{\rho}_{\beta,\zeta}\right)  =k_{B}\left\{  \ln Z_{\zeta
}(J_{\beta})-\sum_{n}N_{n,\zeta}\left(  \beta|\mathrm{out}\right)  [\ln
K_{\zeta}(J_{\beta})]_{nn}\right\}  ,\label{5.21}%
\end{equation}
where $N_{n,\zeta}\left(  \beta|\mathrm{out}\right)  \ $are given by
(\ref{2.36}) with $N_{n,\zeta}(\mathrm{\cdots}|\mathrm{in})=N_{n,\zeta}%
(\beta|\mathrm{in})$.\textrm{ }One can express diagonal elements of $K_{\zeta
}(J_{\beta})$ in terms of the corresponding occupation numbers $N_{n,\zeta
}\left(  \beta|\mathrm{out}\right)  $,%
\begin{equation}
\left[  K_{\zeta}(J_{\beta})\right]  _{nn}=\frac{N_{n,\zeta}\left(
\beta|\mathrm{out}\right)  }{1-\kappa N_{n,\zeta}\left(  \beta|\mathrm{out}%
\right)  },\label{5.22}%
\end{equation}
and do the same to $Z_{\zeta}(J_{\beta})$ by means of the normalization
condition ($\mathrm{tr}_{\zeta}$ $\check{\rho}_{\beta,\zeta}=1$)%
\begin{equation}
Z_{\zeta}(J_{\beta})=\exp\left\{  -\kappa\sum_{n}\ln\left[  1-\kappa
N_{n,\zeta}(\beta|\mathrm{out})\right]  \right\}  ,\label{5.23}%
\end{equation}
to rewrite the expression (\ref{5.21}) for the entropy in the form%
\begin{align}
&  S\left(  \check{\rho}_{\zeta}(\beta)\right)  =\sum_{n}S\left(  \check{\rho
}_{n,\zeta}(\beta)\right)  ,\ \ S\left(  \check{\rho}_{n,\zeta}(\beta)\right)
\nonumber\\
&  \ =-k_{B}\left\{  \kappa\left[  1-\kappa N_{n,\zeta}\left(  \beta
|\mathrm{out}\right)  \right]  \ln\left[  1-\kappa N_{n,\zeta}\left(
\beta|\mathrm{out}\right)  \right]  +N_{n,\zeta}\left(  \beta|\mathrm{out}%
\right)  \ln N_{n,\zeta}\left(  \beta|\mathrm{out}\right)  \right\}
.\label{5.24}%
\end{align}

Considering expressions (\ref{5.24}), (\ref{5.13}), and (\ref{5.4}) one can
see that they all have similar forms.

Next, let us find the Schmidt measure for subsystems of positrons and
electrons for the system with a thermal state as the initial time instant;
subsystems of such a state are described by the reduced density operator
$\check{\rho}_{\zeta}(\beta)$. The entanglement measure of the electron and
positron subsystem is given by
\begin{equation}
\tilde{S}(\check{\rho}_{\zeta}(\beta))=-\mathrm{tr}\text{ }\left[  \check
{\rho}_{\zeta}(\beta)\right]  ^{2}, \label{5.25}%
\end{equation}
where the squares of the operators $\check{\rho}_{\zeta}(\beta)$ are%
\begin{align}
\left[  \check{\rho}_{+}(\beta)\right]  ^{2}  &  =Z_{+}^{-2}(J_{\beta
})\mathbf{:}\exp\left[  \sum\limits_{n}a_{n}^{\dagger}(\mathrm{out})\left[
K_{+}^{2}(J_{\beta})-1\right]  _{nn}a_{n}(\mathrm{out})\right]  \mathbf{:\ ,}%
\nonumber\\
\left[  \check{\rho}_{-}(\beta)\right]  ^{2}  &  =Z_{-}^{-2}(J_{\beta
})\mathbf{:}\exp\left[  \sum\limits_{n}b_{n}^{\dagger}(\mathrm{out})\left[
K_{-}^{2}(J_{\beta})-1\right]  _{nn}b_{n}(\mathrm{out})\right]  \mathbf{:\ ,}
\label{5.26}%
\end{align}
such that%
\begin{equation}
\tilde{S}(\check{\rho}_{\zeta}(\beta))=-\prod_{n}\left\{  1-2\kappa
N_{n,\zeta}\left(  \beta|\mathrm{out}\right)  +(1+\kappa)\left[  N_{n,\zeta
}\left(  \beta|\mathrm{out}\right)  \right]  ^{2}\right\}  ^{\kappa}.
\label{5.27}%
\end{equation}

\subsection{Entropy of measurement-reduced density operators}

The entropy of a density operator $\check{\rho}_{N}$ (\ref{4.8}) has the form%
\begin{equation}
S(\check{\rho}_{N})=-k_{B\text{ }}\mathrm{tr}\text{ }\check{\rho}_{N}\ln
\check{\rho}_{N}. \label{5.28}%
\end{equation}
The representation (\ref{2.21}) allows one to factorize the complete vacuum
into the product of single-mode vacua%
\begin{align}
&  |0,\mathrm{out}\rangle\langle0,\mathrm{out}|\text{ }=\prod\limits_{n}%
|0,\mathrm{out}\rangle_{nn}\langle0,\mathrm{out}|,\text{ \ }\nonumber\\
&  a_{n}(\mathrm{out})|0,\mathrm{out}\rangle_{n}=0,\text{ \ }b_{n}%
(\mathrm{out})|0,\mathrm{out}\rangle_{n}=0. \label{5.29}%
\end{align}
Using this fact and the representation for $|c_{\mathrm{v}}|^{2}$ from
(\ref{4.4}), one can rewrite the density operator (\ref{4.8}) as a product of
single-mode density operators:
\begin{align}
&  \check{\rho}_{N}=%
{\displaystyle\prod\limits_{n}}
\check{\rho}_{N,n},\text{ \ }\mathrm{tr}\check{\rho}_{N,n}=1,\text{ }%
\check{\rho}_{N,n}=|c_{\mathrm{v}}|_{n}^{2}%
{\displaystyle\sum\limits_{f=0}}
W_{f,n}|f,\mathrm{out}\rangle_{nn}\langle f,\mathrm{out}|,\nonumber\\
&  |c_{\mathrm{v}}|_{n}^{2}=|w\left(  -|-\right)  _{nn}|^{-2\kappa},\text{
\ }W_{f,n}=|w\left(  +-|0\right)  _{nn}|^{2f},\text{ \ }|f,\mathrm{out}%
\rangle_{n}=\frac{\left[  a_{n}^{\dagger}(\mathrm{out})b_{n}^{\dagger
}(\mathrm{out})\right]  ^{f}}{f!}|0,\mathrm{out}\rangle_{n}. \label{5.30}%
\end{align}
The quantities $|c_{\mathrm{v}}|_{n}^{2}$ and $|w\left(  +-|0\right)
_{nn}|^{2}$ can be expressed via differential numbers $N_{n}(0|$\textrm{$out$%
}$)$ as
\begin{equation}
|c_{\mathrm{v}}|_{n}^{2}=\left(  1-\kappa N_{n}(0|\mathrm{out})\right)
^{\kappa},\text{ \ }|w\left(  +-|0\right)  _{nn}|^{2}=\frac{N_{n}%
(0|\mathrm{out})}{1-\kappa N_{n}(0|\mathrm{out})}.\text{ \ } \label{5.31}%
\end{equation}
Due to expression (\ref{5.30}), the entropy (\ref{5.28}) can be written as%
\begin{equation}
S(\check{\rho}_{N})=-k_{B\text{ }}%
{\displaystyle\sum\limits_{n}}
\mathrm{tr}\text{ }\check{\rho}_{N,n}\ln\check{\rho}_{N,n}\ . \label{5.32}%
\end{equation}
To calculate the trace of the operator $\check{\rho}_{N,n}\ln\check{\rho
}_{N,n}$, one can use the formal decomposition%
\begin{equation}
\check{\rho}_{N,n}\ln\check{\rho}_{N,n}=\check{\rho}_{N,n}\sum\limits_{k=1}%
^{\infty}k^{-1}\left(  \check{\rho}_{N,n}-1\right)  ^{k}=\sum\limits_{k=1}%
^{\infty}k^{-1}\sum\limits_{l=0}^{k}C_{k}^{l}(\check{\rho}_{N,n}%
)^{l+1}(-1)^{k-l}, \label{5.33}%
\end{equation}
where $C_{k}^{l}$ are binomial coefficients. Due to the orthonormality of the
states\ $|f,\mathrm{out}\rangle_{n}$ the density operators $(\check{\rho
}_{N,n})^{l+1}$ have the form%
\begin{equation}
(\check{\rho}_{N,n})^{l+1}=|c_{\mathrm{v}}|_{n}^{2(l+1)}\sum\limits_{f=0}%
^{\infty}(W_{f,n})^{l+1}|f,\mathrm{out}\rangle_{nn}\langle f,\mathrm{out}|.
\label{5.34}%
\end{equation}
Substituting (\ref{5.34}) into (\ref{5.33}), we obtain
\begin{equation}
\check{\rho}_{N,n}\ln\check{\rho}_{N,n}=|c_{\mathrm{v}}|_{n}^{2}%
\sum\limits_{f=0}^{\infty}W_{f,n}\ln\left(  |c_{\mathrm{v}}|_{n}^{2}%
W_{f,n}\right)  |f,\mathrm{out}\rangle_{nn}\langle f,\mathrm{out}|.
\label{5.35}%
\end{equation}
Then%
\begin{equation}
\mathrm{tr}\text{ }\check{\rho}_{N,n}\ln\check{\rho}_{N,n}=N_{n}%
(0|\mathrm{out})\ln N_{n}(0|\mathrm{out})+\kappa\left[  1-\kappa
N_{n}(0|\mathrm{out})\right]  \ln\left[  1-\kappa N_{n}(0|\mathrm{out}%
)\right]  . \label{5.36}%
\end{equation}
Thus, the entropy of the density operator (\ref{4.8}) reads%
\begin{equation}
S(\check{\rho}_{N})=-k_{B\text{ }}%
{\displaystyle\sum\limits_{n}}
\left\{  \kappa\left[  1-\kappa N_{n}(0|\mathrm{out})\right]  \ln\left[
1-\kappa N_{n}(0|\mathrm{out})\right]  +N_{n}(0|\mathrm{out})\ln
N_{n}(0|\mathrm{out})\right\}  . \label{5.37}%
\end{equation}
The result has the same form as the entropy $S\left(  \check{\rho}_{\zeta
}(0)\right)  $ given by (\ref{5.13}). Thus, we can say that the measurement of
$N$, $N_{+}$ or $N_{-}$ leads to the same information loss as a reduction over
electrons or positrons.

It was shown in the Sec. \ref{S4} that reduction of the density operator
$\check{\rho}_{N}$ over electrons and positrons transforms it in $\left[
\check{\rho}_{N}\right]  _{\zeta}=\check{\rho}_{\zeta}(0)$. This means that if
one calculates the entropy of the density operator $\left[  \check{\rho}%
_{N}\right]  _{\zeta}$, one obtains the same expression (\ref{5.37}) again.
The conditional entropy \cite{NieCh00} $S_{cond}=S(\check{\rho}_{N})-S(\left[
\check{\rho}_{N}\right]  _{\zeta})$, which is used as a measure of
correlations between subsystems, is zero. This fact means that all quantum
correlations between the electrons and positrons are lost due to decoherence,
and there is no entanglement left after the measurement.

\section{T-constant external electric field\label{S6}}

To illustrate some of the above general formulas we consider the so-called
$T$-constant electric field as an external background. Such a field acts only
during a finite time $T$ and it\ is constant within this time interval. Using
this field allows one to avoid troubles with the definition of \textrm{in}-
and \textrm{out}-states inherent to external fields non-switched at
$t\rightarrow\pm\infty$. Another important point is that this field produces a
finite work in a finite space volume. Let us consider $d=(D+1)$-dimensional
space; then the $T$-constant electric field $\mathbf{E}$ is acting during the
time interval $T=t_{\mathrm{out}}-t_{\mathrm{in}}$,%
\begin{equation}
\mathbf{E}=(0,E(t),0,...,0),\,\,E(t)=\left\{
\begin{array}
[c]{l}%
0,\quad-\infty<t\leq t_{\mathrm{in}}\\
E>0,\quad\text{\ }t_{\mathrm{in}}<t<t_{\mathrm{out}}\\
0,\quad\text{\ }\ t_{\mathrm{out}}\leq t<\infty
\end{array}
\right.  , \label{6.1}%
\end{equation}
Processes of pair creation in such a field were studied in Refs.
\cite{23,26,GavGi96,GavGi08}). Similar to these works, we consider
sufficiently large $T$.

Since there is no particle production after the time instant $t_{\mathrm{out}%
}$, differential mean numbers of particles\ $N_{n,\zeta}(\mathrm{\cdots}%
|$\textrm{$out$}$)$ created in a given state $n=\mathbf{p},r$ ($\mathbf{p}$ is
a $D$-dimensional vector of momentum and $r$ is spin) depend only on the time
interval. The electric field acting during the sufficiently long time $T$
creates a considerable number of pairs only in a finite region in the momentum
space. Since we suppose $T\gg\max\{1,E_{c}/E\}$, we need to consider only the
range%
\begin{equation}
|p_{\bot}|\leq\sqrt{eE}\left[  \sqrt{eE}T\right]  ^{1/2},\ \ -T/2\leq
p_{1}/eE\leq T/2 \label{6.2}%
\end{equation}
in the momentum space, see \cite{GavGi96} for details. Note that for the case
$d=2$ there are no transversal components of momentum.

\subsection{Vacuum initial state}

First let us consider the case when the system initially was in the vacuum
state. For this case the differential mean numbers in the momentum range
(\ref{6.2}) are
\begin{equation}
N_{n}\left(  0|\mathrm{out}\right)  =e^{-\pi\lambda},\ \ \lambda=(p_{\bot}%
^{2}+m^{2})/eE\text{ }. \label{a}%
\end{equation}
They have the same form as in the case of the constant uniform electric field
\cite{FroGi78, Nik69} and are the same for bosons and fermions. The entropy
(\ref{5.13}) is expressed in terms of $N_{n}\left(  \mathrm{0}|\mathrm{out}%
\right)  $ and does not depend on the spin quantum number $r$, thus, the
summation over the latter results in the factor $\gamma_{(d)}=2^{[\frac{d}%
{2}]-1}$.

First we consider the Dirac case with $\kappa=+1$:
\begin{equation}
S(\check{\rho}_{n,\zeta}(0))=-k_{B}\left\{  \left[  1-N_{n}\left(
0|\mathrm{out}\right)  \right]  \ln\left[  1-N_{n}\left(  0|\mathrm{out}%
\right)  \right]  +N_{n}\left(  0|\mathrm{out}\right)  \ln N_{n}\left(
0|\mathrm{out}\right)  \right\}  . \label{6.8}%
\end{equation}
For the case of electric field the mean number of particles created
$N_{n}\left(  0|\mathrm{out}\right)  $ can vary only within the range $(0,1)$
and depends only on thestrength of the external field. Expression (\ref{6.8})
is symmetric with respect to $N_{n,\zeta}\left(  0|\mathrm{out}\right)  $. It
reaches maximum at $N_{n}\left(  0|\mathrm{out}\right)  =1/2$ and turns to
zero at $N_{n}\left(  0|\mathrm{out}\right)  =1$ and $N_{n}\left(
0|\mathrm{out}\right)  =0$. This fact can be interpreted as follows. In the
case of $N_{n}\left(  0|\mathrm{out}\right)  =0$ there are no particles
created by the external field and theinitial vacuum state in the mode remains
unchanged. The case $N_{n}\left(  0|\mathrm{out}\right)  =1$ corresponds to
the situation when a particle is created with certainty. The maximum of
(\ref{6.8}), corresponding to $N_{n}\left(  0|\mathrm{out}\right)  =1/2$, is
associated with the state with the maximum amount of uncertainty.

Representing the logarithm in the first term of expression (\ref{6.8}) as the
Taylor series in powers of $N_{n}\left(  0|\mathrm{out}\right)  $, we see that
$S(\check{\rho}_{n,\zeta}(0))$ is proportional to $N_{n}\left(  0|\mathrm{out}%
\right)  $. The latter plays the role of the cut-off parameter for the
integral over $p_{1}$ \cite{GavGi96}. Thus, the summation over the quantum
numbers can be reduced to an integration over momenta that satisfy
restrictions (\ref{6.2}),%
\[
\sum_{n}\rightarrow\frac{\gamma_{(d)}V}{\left(  2\pi\right)  ^{d-1}}\int
d\mathbf{p,}%
\]
where $V$ is the $D$-dimensional spatial volume. The mean numbers (\ref{a}) do
not depend on the longitudinal component of momentum. Outside of the range
(\ref{6.2}), the contribution to the integral is very small, and this allows
us to extend the integration limits of $p_{\bot}$ to infinity. Integration
over $p_{\bot}$ can be performed using the Taylor series. The result of the
integration is
\begin{equation}
S(\check{\rho}_{\zeta}(0))=\gamma_{(d)}k_{B}\frac{\left(  eE\right)
^{\frac{d}{2}}TV}{(2\pi)^{d-1}}A_{\mathrm{Dirac}}\left(  d,E_{c}/E\right)  ,
\label{6.10}%
\end{equation}
where the factor $TV$ can be considered as the $d$-dimensional volume. To get
finite and correct expressions, one should use the volume normalization. The
factor $A_{\mathrm{Dirac}}\left(  d,E_{c}/E\right)  $ has the form
\begin{align*}
&  A_{\mathrm{Dirac}}\left(  d,E_{c}/E\right)  =\sum\limits_{l=1}^{\infty
}l^{-d/2}\exp[-\pi lE_{c}/E]\\
&  -\sum\limits_{l=1}^{\infty}l^{-1}(l+1)^{\left(  2-d\right)  /2}\exp
[-\pi(l+1)E_{c}/E]+\left(  \pi\frac{E_{c}}{E}+\frac{d-2}{2}\right)
\exp\left(  -\pi E_{c}/E\right)  .
\end{align*}

It is possible to estimate the entropy in strong-field $E_{c}/E\ll1$,
critical-field $E_{c}/E=1$, and weak-feald $E_{c}/E\gg1$ limits. For example,
for a strong field with $d=4$ we have $A_{\mathrm{Dirac}}\left(  4,0\right)
=\pi^{2}/6$, for the critical field, we have $A_{\mathrm{Dirac}}\left(
4,1\right)  \approx0,22$. In the case of a weak field the entropy has a small
value of the order of $(\pi E_{c}/E)\exp[-\pi E_{c}/E]$ for any $d$. For $d=3$
the following estimations hold $A_{\mathrm{Dirac}}\left(  3,0\right)
\approx0,93$, $A_{\mathrm{Dirac}}\left(  3,1\right)  \approx0,2$;\emph{ }for
$d=2$ the factor $A\left(  2,0\right)  $ is a value of order of $1$, and
$A\left(  2,1\right)  =e^{-\pi}$.

Let us consider the KG case ($\kappa=-1$),
\begin{equation}
S(\check{\rho}_{n,\zeta}(0))=k_{B}\left\{  \left[  1+N_{n}\left(
0|\mathrm{out}\right)  \right]  \ln\left[  1+N_{n}\left(  0|\mathrm{out}%
\right)  \right]  -N_{n}\left(  0|\mathrm{out}\right)  \ln N_{n}\left(
0|\mathrm{out}\right)  \right\}  . \label{6.12}%
\end{equation}
Expression (\ref{6.12}) just increases with $N_{n}\left(  0|\mathrm{out}%
\right)  $. After summation over the quantum numbers, the entropy (\ref{5.11})
takes the form
\begin{equation}
S(\check{\rho}_{\zeta}(0))=k_{B}\frac{\left(  eE\right)  d^{d/2}TV}%
{(2\pi)^{d-1}}A_{\mathrm{KG}}\left(  d,E_{c}/E\right)  , \label{6.13}%
\end{equation}
where
\begin{align*}
&  A_{\mathrm{KG}}\left(  d,E_{c}/E\right)  =\sum\limits_{l=1}^{\infty
}l^{-d/2}(-1)^{l-1}\exp[-\pi lE_{c}/E]\\
&  +\sum\limits_{l=1}^{\infty}l^{-1}(l+1)^{\left(  2-d\right)  /2}%
(-1)^{l-1}\exp[-\pi(l+1)E_{c}/E]+\left(  \pi\frac{E_{c}}{E}+\frac{d-2}%
{2}\right)  \exp\left(  -\pi E_{c}/E\right)  .
\end{align*}

The following are estimations for different field strengths: $A_{\mathrm{KG}%
}\left(  4,0\right)  \approx2,21$, $\ A_{\mathrm{KG}}\left(  4,1\right)
\approx0,22$; $A_{\mathrm{KG}}\left(  3,0\right)  \approx1,78$,
$A_{\mathrm{KG}}\left(  3,1\right)  \approx0,2$; $A_{\mathrm{KG}}\left(
2,0\right)  \approx1$, $A_{\mathrm{KG}}\left(  2,1\right)  \approx e^{-\pi}$.
In the case of weak field the entropy is a small value of the order of $(\pi
E_{c}/E)\exp[-\pi E_{c}/E]$ for any $d$ again.

We have mentioned before that the entropy of the density operator $\check
{\rho}_{N}$, given by (\ref{5.37}), is exactly of the same form as the entropy
of $\check{\rho}_{\zeta}(0)$, given by (\ref{5.13}), hence all the
considerations for the case with the intermediate measurement are the same.

\subsection{Mixed initial state}

We note that the entropy (\ref{5.24}) of the system that has been in thermal
equilibrium at the initial time instant\textrm{ }is expressed in terms of
differential mean numbers $N_{n,\zeta}\left(  \beta|\mathrm{out}\right)  $
(\ref{2.36}) of particles created by the external field, whereas initial
differential numbers of particles are%
\begin{equation}
N_{n,\zeta}\left(  \beta|\mathrm{in}\right)  =\left[  \exp\beta(\varepsilon
_{n}-\mu)+\kappa\right]  ^{-1},\text{ \ }\varepsilon_{n}=\sqrt{m^{2}+p_{\bot
}^{2}+(p_{1}+eET/2)^{2}}. \label{6.16}%
\end{equation}
Let us discuss two cases, the first one being the case of low temperature%
\[
\beta(\epsilon_{\bot}-\mu)\gg1,\ \ \epsilon_{\bot}=\sqrt{m^{2}+p_{\bot}^{2}},
\]
when all the energies of the particles created with a given $p_{_{\bot}}$ are
considerably higher than the temperature, and the second being the case of
high temperature $\beta eET\ll1$, when all the energies of the created
particles are much lower than the temperature. We assume for simplicity that
$eET$ $\gg\mu$ and that$T$ is sufficiently large to provide $\left(
eET\right)  ^{2}\gg m^{2}+p_{\bot}^{2}$.

In the low-temperature case, the number of particles created does not depend
on the longitudinal momenta:%
\[
N_{n,\zeta}\left(  \beta|\mathrm{in}\right)  \approx\exp\left(  -\beta
\varepsilon_{n}\right)  \rightarrow0,\ \ N_{n,\zeta}\left(  \beta
|\mathrm{out}\right)  \rightarrow N_{n,\zeta}\left(  0|\mathrm{out}\right)  .
\]
In this limit entropy $S(\check{\rho}_{n,\zeta}(\beta))$ tends to that of the
zero-temperature case (initial vacuum state). Then integration over
transversal momenta can be done exactly as in the initial vacuum case.

Formal calculations of $N_{n,\zeta}\left(  \beta|\mathrm{out}\right)  $ and of
the entropy in the case of high temperature, $\beta eET\ll1$, are also quite
simple. However, it was shown in Ref. \cite{GavGi08} that in the Dirac case
under such a condition the current density is much greater than the current
density of particles created from the vacuum, due to the work of the external
field performed over the particles [which was denoted by $\operatorname{Re}%
\langle j_{\mu}\left(  t\right)  \rangle_{\theta}^{c}$ in \cite{GavGi08}]
already existing in the initial state. Therefore, in such a case the particle
creation effect may be disregarded.

We note that the general form of the reduced density operators $\check{\rho
}_{\pm}$ , given by Eq.\ (\ref{3.3}), allows one to study the change of the
entropy and the corresponding entanglement during many consecutive
measurements. In this case the density operator $\check{\rho}_{N}$
(\ref{4.8b}) has to be considered as the initial state for the second stage of
the evolution and so on. In the general case it is not simple to describe such
a decoherence procedure for an arbitrary stage. However, as was mentioned
above, if the mean numbers $N_{n}(0|\mathrm{out})$ are not small within a
sufficiently large range of momenta, already on the second stage the
particle-creation effect may be disregarded and the subsequent decoherence is
described in the usual terms.

\section{Summary}

Using a general nonperturbative expression for the density operators (of
quantized Dirac or KG fields), we derived their specific forms corresponding
to different initial conditions. Applying a reduction procedure to specific
density operators, we constructed mixed states of both electron and positron
subsystems. Calculating the entropy of such states, we obtained the loss of
information due to the reduction and, at the same time, the entanglement of
electron and positron subsystems. We paid attention to the fact that any
measurement in the system under consideration implies a decoherence and the
corresponding modifications of the complete and the reduced density operators.
We studied the results of such a decoherence and we related to it the loss of
information by calculating the information entropy. To illustrate some of the
obtained general results, we considered the slowly varying $T$-constant
electric field as an external background. We derived the following
conclusions. The entropy of any subsystem (of electrons or positrons) with the
vacuum as the initial state is proportional to the factor $\left(  eE\right)
^{d/2}$ and to the number of spin degrees of freedom $\gamma_{(d)}$. It grows
linearly with the time of the field action $T$. The above behavior remains in
the thermal case at low temperatures; in fact, here the entropy does not
depend on the temperature.

{\Large Acknowledgements}

SPG thanks FAPESP for support and University of São Paulo for hospitality. DMG
is grateful to the Brazilian foundations FAPESP and CNPq for support. The work
of SPG and DMG was also partially supported by the Tomsk State University
Competitiveness Improvement Program. AAS thanks CAPES for support. The
reported study of SPG, DMG, and AAS was partially supported by RFBR Project
No. 15-02-00293a.

\appendix

\section{QED with strong electric-like background\label{Ap}}

In this appendix, we consider briefly a special case of QFT with an unstable
vacuum of the quantized Dirac or KG field with time-dependent electriclike
background that is switched on and off at $t\rightarrow\pm\infty$.
Quantization of this theory in terms of \textrm{in}- and \textrm{out}%
-electrons and positrons was elaborated in Refs. \cite{Gitman}. Some results
of this quantization necessary for us here\emph{ }are presented below.

We denote operators in the Schrödinger representation by a caret, e.g.,
$\hat{A}$, while operators in the Heisenberg representation are denoted by an
inverted caret, g.e., $\check{A}$.\emph{ }In the Schrödinger picture one can
define the following: at the initial time instant $t_{\mathrm{in}}$, a set of
creation and annihilation operators $a_{n}^{\dagger}(t_{\mathrm{in}})$ and
$a_{n}(t_{\mathrm{in}})$ of electrons and similar operators $b_{n}^{\dagger
}(t_{\mathrm{in}})$ and $b_{n}(t_{\mathrm{in}})$ of positrons such that the
corresponding vacuum at $t_{\mathrm{in}}$ is $|0,t_{\mathrm{in}}\rangle$; at
the final time instant $t_{\mathrm{out}}$, a set of creation and annihilation
operators $a_{n}^{\dagger}(t_{\mathrm{out}})$, $a_{n}(t_{\mathrm{out}})$, of
electrons and similar operators $b_{n}^{\dagger}(t_{\mathrm{out}})$,
$b_{n}(t_{\mathrm{out}})$ of positrons, such that the corresponding vacuum at
$t_{\mathrm{out}}$ is $|0,t_{\mathrm{out}}\rangle$,%
\[
a_{n}(t_{\mathrm{in}})|0,t_{\mathrm{in}}\rangle\;=b_{n}(t_{\mathrm{in}%
})|0,t_{\mathrm{in}}\rangle=0\,,\;a_{n}(t_{\mathrm{out}})|0,t_{out}%
\rangle\;=b_{n}(t_{\mathrm{out}})|0,t_{\mathrm{out}}\rangle=0\,\;\;\forall n.
\]
The probability amplitude for the transition from an initial state to a final
state $M_{\mathrm{in}\rightarrow\mathrm{out}}$ has the following form in the
Schrödinger picture:
\[
M_{\mathrm{in}\rightarrow\mathrm{out}}=\langle t_{\mathrm{out}}%
|U(t_{\mathrm{out}},t_{\mathrm{in}})|t_{\mathrm{in}}\rangle,
\]
where $U(t,t^{\prime})$ is a unitary evolution operator of the system. The
density operator of an initial state $\hat{\rho}\left(  t_{\mathrm{in}%
}\right)  $ is given as an operator-valued function of the creation and
annihilation operators of electrons (positrons) at the initial time instant%
\[
\hat{\rho}\left(  t_{\mathrm{in}}\right)  =\rho_{\mathrm{in}}\left(
a^{\dagger}(t_{\mathrm{in}}),a(t_{\mathrm{in}}),b^{\dagger}(t_{\mathrm{in}%
}),b(t_{\mathrm{in}})\right)  .
\]

The mean value of a physical quantity $F$ at the final time instant reads%
\begin{equation}
\langle F(t_{\mathrm{out}})\rangle=\mathrm{tr}\text{ }\hat{\rho}%
(t_{\mathrm{out}})\hat{F}(t_{\mathrm{out}}), \label{a1}%
\end{equation}
where $\hat{\rho}(t)$ is the density operator in the Schrödinger
representation at time instant $t$, and the designation $\mathrm{tr}$ stands
for the complete trace,%
\begin{equation}
\hat{\rho}(t_{\mathrm{out}})=U\left(  t_{\mathrm{out}},t_{\mathrm{in}}\right)
\hat{\rho}\left(  t_{\mathrm{in}}\right)  U^{\dagger}\left(  t_{\mathrm{out}%
},t_{\mathrm{in}}\right)  \,. \label{a2}%
\end{equation}

In order to pass to the Heisenberg picture we define the finite-time evolution
operators $\Omega_{\left(  \pm\right)  }$,
\begin{align}
&  \Omega_{\left(  +\right)  }=U\left(  0,t_{\mathrm{in}}\right)
\,,\;\Omega_{\left(  -\right)  }=U\left(  0,t_{\mathrm{out}}\right)
\,,\;U\left(  t_{\mathrm{out}},t_{\mathrm{in}}\right)  =\Omega_{\left(
-\right)  }^{\dagger}\Omega_{\left(  +\right)  }\,,\nonumber\\
&  \check{\rho}=\hat{\rho}(0)=\Omega_{\left(  +\right)  }\hat{\rho}\left(
t_{\mathrm{in}}\right)  \Omega_{(+)}^{\dagger}=\Omega_{(-)}\hat{\rho}\left(
t_{\mathrm{out}}\right)  \Omega_{(-)}^{\dagger}\ ; \label{a4}%
\end{align}
a set of creation and annihilation operators $a_{n}^{\dagger}(\mathrm{in})$
and $a_{n}(\mathrm{in})$ of \textrm{in}-electrons, similar operators
$b_{n}^{\dagger}(\mathrm{in})$ and $b_{n}(\mathrm{in})$ of \textrm{in}
positrons, and the corresponding \textrm{in} vacuum $|0,\mathrm{in}\rangle$;
and a set of creation and annihilation operators $a_{n}^{\dagger}%
(\mathrm{out})$, $a_{n}(\mathrm{out})$, of \textrm{out} electrons, similar
operators $b_{n}^{\dagger}(\mathrm{out})$ and $b_{n}(\mathrm{out})$ of
\textrm{out} positrons, and the corresponding \textrm{out} vacuum
$|0,\mathrm{out}\rangle$,
\begin{align}
&  \left\{  a(\mathrm{in}),\ldots\right\}  =\Omega_{(+)}\left\{
a(t_{\mathrm{in}}),\cdots\right\}  \Omega_{(+)}^{\dagger},\ |0,\mathrm{in}%
\rangle=\Omega_{(+)}|0,t_{\mathrm{in}}\rangle\,,\nonumber\\
&  \left\{  a\left(  \mathrm{out}\right)  ,\ldots\right\}  =\Omega
_{(-)}\left\{  a(t_{\mathrm{out}}),\ldots\right\}  \Omega_{(-)}^{\dagger
},\;|0,\mathrm{out}\rangle=\Omega_{(-)}|0,t_{\mathrm{out}}\rangle
\,,\nonumber\\
&  M_{\mathrm{in}\rightarrow\mathrm{out}}=\langle0,t_{\mathrm{out}}|\cdots
a(t_{\mathrm{in}})\Omega_{\left(  -\right)  }^{\dagger}\Omega_{\left(
+\right)  }a_{n}^{\dagger}(t_{\mathrm{in}})\cdots|0,t_{\mathrm{in}}%
\rangle=\langle0,\mathrm{out}|\cdots a(\mathrm{out})a_{n}^{\dagger
}(\mathrm{in})\cdots|0,\mathrm{in}\rangle,\nonumber\\
&  c_{v}=\langle0,t_{\mathrm{out}}|U\left(  t_{\mathrm{out}},t_{\mathrm{in}%
}\right)  |0,t_{\mathrm{in}}\rangle=\langle0,\mathrm{out}|0,\mathrm{in}%
\rangle. \label{a5}%
\end{align}

All the information concerning the processes of particle creation,
annihilation and scattering is contained in the elementary probability
amplitudes
\begin{align}
&  w\left(  +|+\right)  _{mn}=c_{v}^{-1}\langle0,\mathrm{out}\left\vert
a_{m}(\mathrm{out})a_{n}^{\dagger}(\mathrm{in})\right\vert 0,\mathrm{in}%
\rangle,\nonumber\\
&  w\left(  -|-\right)  _{nm}=c_{v}^{-1}\langle0,\mathrm{out}\left\vert
b_{m}(\mathrm{out})b_{n}^{\dagger}(\mathrm{in})\right\vert 0,\mathrm{in}%
\rangle\,,\nonumber\\
&  w\left(  0|-+\right)  _{nm}=c_{v}^{-1}\langle0,\mathrm{out}\left\vert
b_{n}^{\dagger}(\mathrm{in})a_{m}^{\dagger}(\mathrm{in})\right\vert
0,\mathrm{in}\rangle\,,\nonumber\\
&  w\left(  +-|0\right)  _{mn}=c_{v}^{-1}\langle0,\mathrm{out}\left\vert
a_{m}(\mathrm{out})b_{n}(\mathrm{out})\right\vert 0,\mathrm{in}\rangle\,.
\label{a6}%
\end{align}

The amplitudes (\ref{a6}) can be calculated with the help of certain
appropriate sets of solutions of the corresponding relativistic wave equation
with an external field (Klein-Gordon, Dirac, and so on) (see \cite{Gitman}).
We are interested in the case of a uniform external field, which does not mix
different quantum modes. Thus, in this paper the amplitudes (\ref{a6}) are
diagonal in quantum numbers,
\[
w\left(  \zeta|\zeta\right)  _{mn}=\delta_{mn}w\left(  \zeta|\zeta\right)
_{nn},\ w\left(  0|-+\right)  _{nm}=\delta_{mn}w\left(  0|-+\right)
_{nn},\ w\left(  +-|0\right)  _{nm}=\delta_{mn}w\left(  +-|0\right)  _{nn}.
\]

The sets of \textrm{in}- and \textrm{out}-operators are related to each other
by a linear canonical transformation \cite{Ber65}, which can be written in
terms of the amplitudes (\ref{a6}),\footnote{We use condensed notation, for
example,
\[
bw\left(  0|-+\right)  a=\sum_{n,m}b_{n}w\left(  0|-+\right)  _{nm}a_{m}\,.
\]
}
\begin{align}
&  a(\mathrm{out})=\left[  w\left(  +|+\right)  ^{\dagger}\right]
^{-1}a(\mathrm{in})-\kappa w\left(  +-|0\right)  \left[  w\left(  -|-\right)
\right]  ^{-1}b^{\dagger}(\mathrm{in}),\nonumber\\
&  b^{\dagger}(\mathrm{out})=\left[  w\left(  +|+\right)  ^{\dagger}\right]
^{-1}w\left(  +-|0\right)  ^{\dagger}a(\mathrm{in})+\left[  w\left(
-|-\right)  \right]  ^{-1}b^{\dagger}(\mathrm{in}), \label{a7}%
\end{align}
and by its Hermitian conjugate. As it has been demonstrated \cite{Gitman}),
such a relation is given by a unitary operator $V$,
\begin{equation}
V\left\{  a\left(  \mathrm{out}\right)  ,\ldots\right\}  V^{\dagger}=\left\{
a\left(  \mathrm{in}\right)  ,\ldots\right\}  ,\ \ \,|0,\mathrm{in}%
\rangle=V|0,\mathrm{out}\rangle\,, \label{a8}%
\end{equation}
which has the form $V=v_{4}v_{3}v_{2}v_{1}$,%
\begin{align}
&  v_{1}=\exp\left\{  -\kappa b(\mathrm{out})w\left(  0|-+\right)
a(\mathrm{out})\right\}  \,,\;v_{2}=\exp\left\{  a^{\dagger}(\mathrm{out})\ln
w\left(  +|+\right)  a(\mathrm{out})\right\}  \,,\nonumber\\
&  v_{3}=\exp\left\{  -\kappa b(\mathrm{out})\ln w\left(  -|-\right)
b^{\dagger}(\mathrm{out})\right\}  \,,\;v_{4}=\exp\left\{  -\kappa a^{\dagger
}(\mathrm{out})w\left(  +-|0\right)  b^{\dagger}(\mathrm{out})\right\}  \,.
\label{a9}%
\end{align}
Using this expression for $V$, one can find
\begin{equation}
c_{v}=\langle0,\mathrm{out}|V|0,\mathrm{out}\rangle=\exp\left\{
-\kappa\mathrm{tr}\ln w\left(  -|-\right)  \right\}  \,. \label{a10}%
\end{equation}

\end{document}